\documentclass[letterpaper]{article} 
\usepackage{aaai2026}
\usepackage{times}  
\usepackage{helvet}  
\usepackage{courier}  
\usepackage[hyphens]{url}  
\usepackage{graphicx} 
\usepackage{natbib}  
\usepackage{caption} 
\usepackage{algorithm}
\usepackage{algorithmic}
\usepackage{amsfonts}
\usepackage{amsmath}
\usepackage{amssymb}

\usepackage{newfloat}
\usepackage{listings}
\DeclareCaptionStyle{ruled}{labelfont=normalfont,labelsep=colon,strut=off} 
\floatstyle{ruled}
\newfloat{listing}{tb}{lst}{}
\floatname{listing}{Listing}
\usepackage{tcolorbox}
\usepackage{makecell}

\title{AutoMalDesc: Large-Scale Script Analysis for Cyber Threat Research}
\author{
    Alexandru-Mihai Apostu\textsuperscript{\rm 1,2},
    Andrei Preda\textsuperscript{\rm 1},
    Alexandra Daniela Damir\textsuperscript{\rm 1},
    Diana Bolocan\textsuperscript{\rm 1},\\
    Radu Tudor Ionescu\textsuperscript{\rm 2},
    Ioana Croitoru\textsuperscript{\rm 1},
    Mihaela Gaman\textsuperscript{\rm 1}
}
\affiliations{
    \textsuperscript{\rm 1}CrowdStrike\\
    \textsuperscript{\rm 2}University of Bucharest, Romania\\
}

\usepackage{bibentry}

\begin{document}

\maketitle

\begin{abstract}
Generating thorough natural language explanations for threat detections remains an open problem in cybersecurity research, despite significant advances in automated malware detection systems. In this work, we present AutoMalDesc, an automated static analysis summarization framework that, following initial training on a small set of expert-curated examples, operates independently at scale. This approach leverages an iterative self-paced learning pipeline to progressively enhance output quality through synthetic data generation and validation cycles, eliminating the need for extensive manual data annotation. Evaluation across 3,600 diverse samples in five scripting languages demonstrates statistically significant improvements between iterations, showing consistent gains in both summary quality and classification accuracy. Our comprehensive validation approach combines quantitative metrics based on established malware labels with qualitative assessment from both human experts and LLM-based judges, confirming both technical precision and linguistic coherence of generated summaries. To facilitate reproducibility and advance research in this domain, we publish our complete dataset of more than 100K script samples, including annotated seed (0.9K) and test (3.6K) datasets, along with our methodology and evaluation framework.

\end{abstract}

\begin{links}
    \link{Code}{https://github.com/CrowdStrike/automaldesc}
\end{links}


\section{Introduction}
\label{sect:intro}

\begin{figure}[h!]
\centering
\includegraphics[width=0.85\columnwidth]{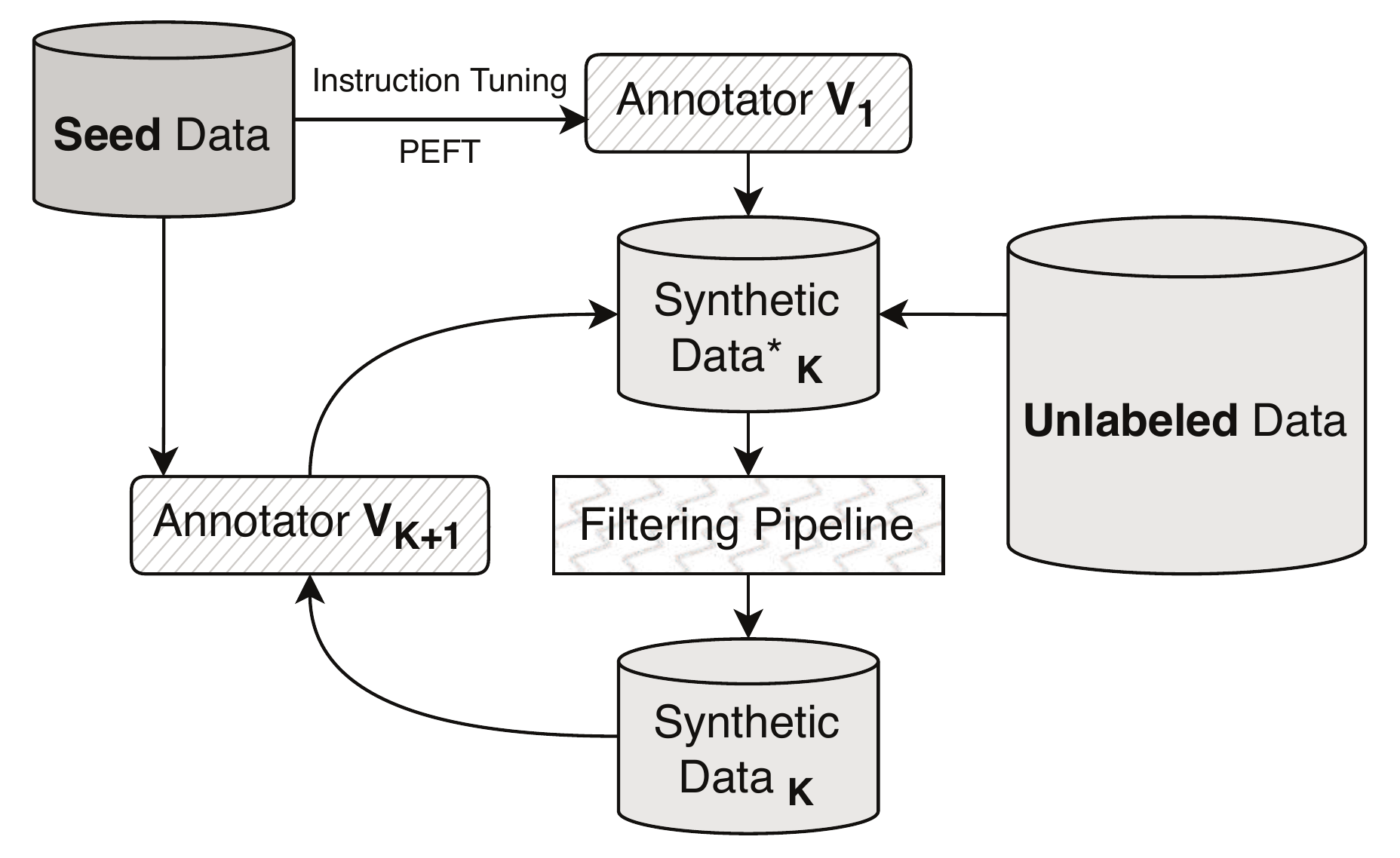}
\caption{Self-training methodology. A \textit{seed dataset} of 900 high-quality labeled samples initiates semi-supervised learning through pseudo-labeling. An LLM trained on seed data generates filtered pseudo-labels for unlabeled data, enabling iterative improvement with expanding training data.}
\label{fig:methodology}
\end{figure}

\begin{figure}[h!]
\centering
\includegraphics[width=0.99\columnwidth]{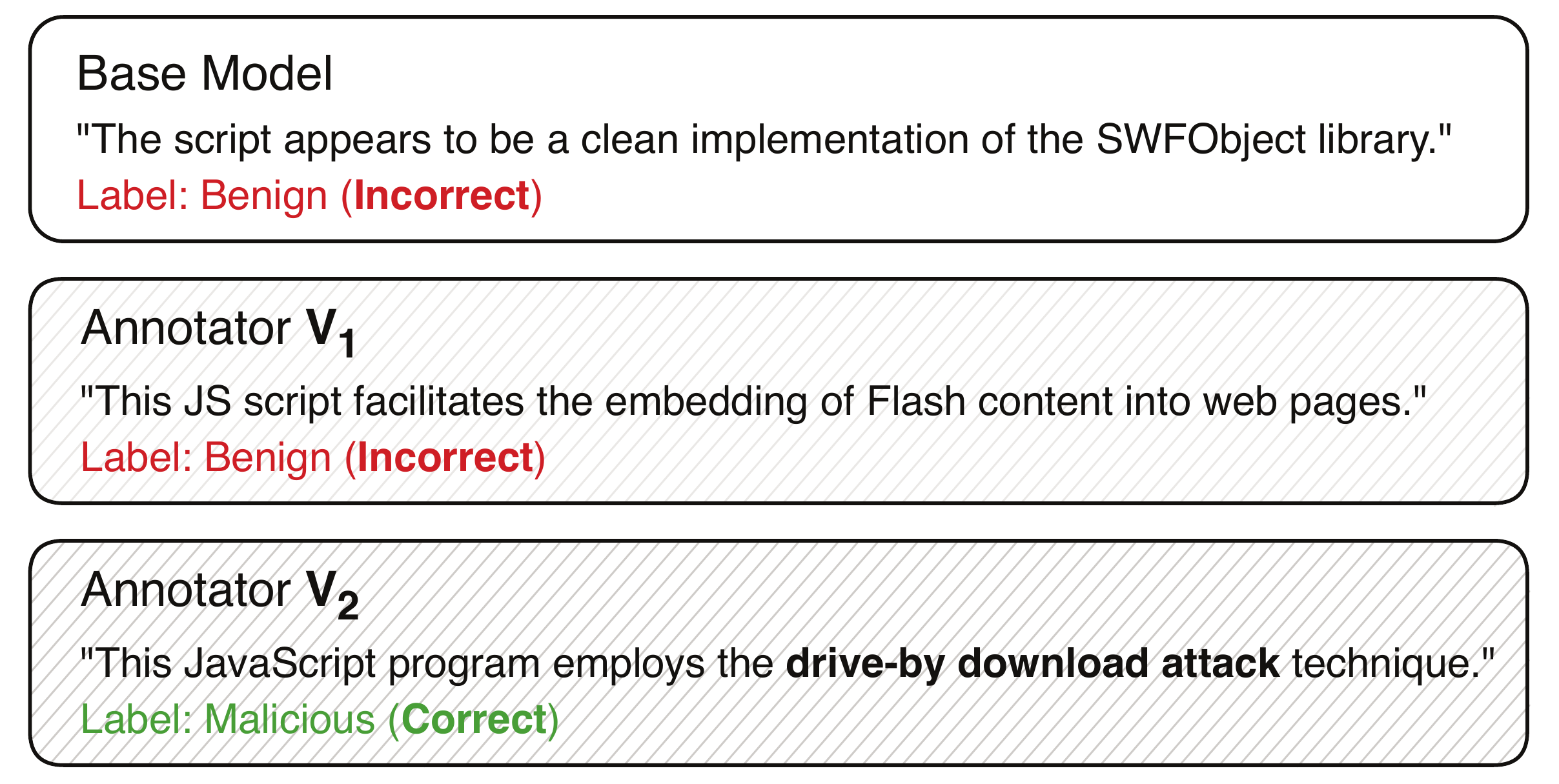}
\caption{Evolution of model predictions for a malicious JavaScript program. The pretrained (base) model is unable to recognize the malicious intent of the code, focusing on surface-level details. Through self-learning, the model learns to detect malware and identify the attack mechanism.}
\label{fig:examples}
\end{figure}

\begin{table*}[t]
\centering
\begin{tabular}{l|rrr|rrr|rrr|rrr}
    & \multicolumn{3}{c|}{Initial Corpus$^*$} & \multicolumn{3}{c|}{Training Data$^*$} & \multicolumn{3}{c|}{Test Set} & \multicolumn{3}{c}{Seed Dataset} \\
    Lang & Benign & Mal. & Total & Benign & Mal. & Total & Benign & Mal. & Total & Benign & Mal. & Total \\
    \hline
    sh        & 16,207 & 20,616 & 36,823 & 9,059 & 19,926 & 28,985 & 295 & 512 & 807 & 66 & 100 & 166 \\
    bat       & 5,000 & 2,838 & 7,838 & 2,775 & 2,575 & 5,350 & 594 & 226 & 820 & 64 & 103 & 167 \\
    js        & 35,000 & 30,000 & 65,000 & 21,242 & 14,132 & 35,374 & 380 & 475 & 855 & 95 & 139 & 234 \\
    ps        & 12,827 & 17,007 & 29,834 & 9,337 & 6,820 & 16,157 & 311 & 345 & 656 & 61 & 105 & 166 \\
    py        & 10,236 & 7,395 & 17,631 & 7,584 & 7,827 & 15,411 & 236 & 262 & 498 & 66 & 101 & 167 \\
    \hline
    all       & 79,270 & 77,856 & 157,126 & 49,997 & 51,280 & 101,277 & 1,816 & 1,820 & 3,636 & 352 & 548 & 900 \\
\end{tabular}
\caption{Distribution of scripts across datasets by language and maliciousness labels. The seed and test sets underwent rigorous validation through multiple sources, including expert verification. $^*$Initial corpus and training set labels are derived from internal labeling automation and shown for reference only. Due to the inherent noise of these labels, the training set was treated as unlabeled data. Languages: Bash (sh), Batch (bat), JavaScript (js), PowerShell (ps) and Python (py).}
\label{tab_data_distrib}
\end{table*}

As large language models continue to excel across diverse domains~\cite{Roberts-IEEE-CVPRW-2024,Wang-IEEE-2024,Thirunavukarasu-Nature-2023,Wang-Nature-2023,Rivas-AI-2023}, their increasing data demands has sparked significant interest into self-rewarding and self-improving approaches~\cite{Dong-ICLR-2025,Zelikman-NeurIPS-2024,Liang-CoRR-2024,Huang-EMNLP-2023,Gulcehre-ArXiv-2023}. This is particularly relevant in specialized fields where expert-labeled data is scarce and expensive to obtain. 

The cybersecurity sector exemplifies these challenges, where rapidly evolving threats and restricted access to malware samples create substantial barriers to dataset creation~\cite{Guo-ASE-2023}. The dynamic threat landscape, combined with scarce expert annotations and classified data, makes self-improvement techniques essential for advancing security-focused LLM applications. Beyond the data constraints, another challenge lies in generating detailed, human-readable descriptions of malware behavior. While current analysis tools excel at threat detection and classification~\cite{Wang-CNCC-2025,Rudd-DigitalThreats-2024,Dambra-CCS-2023,Eren-ACM-2023,Wu-ICCNS-2018}, they fall short in automatically producing comprehensive behavioral descriptions~\cite{He-ISSTA-2025}. This gap between technical detection capabilities and the need for interpretable analysis represents a key bottleneck in current research~\cite{He-ISSTA-2025,Jelodar-ArXiv-2025,Fujii-ESORICS-2024}.

To bridge these disparities, we present the first comprehensive study on self-improving LLMs for malware script analysis, with three key contributions:
\begin{itemize}
    \item \textbf{Self-improving pipeline:} a novel iterative methodology for enhancing cybersecurity script analysis through LLM-generated annotations, with empirical validation of its effectiveness (see Figures \ref{fig:methodology} and \ref{fig:examples}).
    
    \item \textbf{Public dataset:} a balanced collection of 157K scripts (78K malicious, 79K benign), spanning 5 programming languages relevant for malware analysis (see Table \ref{tab_data_distrib}).
    
    \item \textbf{Evaluation framework:} a comprehensive benchmark combining quantitative metrics (maliciousness label and language accuracy) with human and LLM-Judge assessments of technical precision and summary quality.
\end{itemize}


\section{Related Work}
Our research builds upon two key areas of prior work: self-improving language models that can enhance their capabilities through iterative learning, and static security analysis techniques for script-based threats. 
\label{sect:related}

\subsection{Iterative and Self-Improving Approaches} 
Prior research shows that language models can enhance their capabilities through self-generated training data. Earlier work by \citet{Wang-ArXiv-2022} introduced Self-Instruct, bootstrapping from a small set of human prompts to generate and filter instruction-response pairs for model fine-tuning. Building on this concept, the STaR system \citep{Zelikman-NeurIPS-2024} implemented a feedback loop where models generate step-by-step reasoning and learn from successful examples, significantly improving performance on mathematical and commonsense tasks. 
 Several strategies have emerged to tackle self-improvement with limited ground-truth data. \citet{Huang-EMNLP-2023} leveraged Chain-of-Thought prompting and majority voting across multiple reasoning paths to identify reliable predictions, while \citet{Li-ICLR-2024} developed instruction backtranslation, by deriving prompts from web documents. For resource-constrained settings, \citet{Dong-ICLR-2025} proposed Self-Boost, which uses a single model to generate and verify new examples. Recent innovations include \citet{Yuan-ICML-2024}'s self-rewarding models using LLM-based judgment, \citet{Gulcehre-ArXiv-2023}'s reinforcement learning strategy, combining data generation with offline fine-tuning, and \citet{Wang-ArXiv-2024}'s extension of self-improvement to evaluation through model-generated comparisons.

Our work adapts these self-improvement techniques to cybersecurity, specifically addressing the critical challenge of data scarcity in malware analysis.

\subsection{Static Security Analysis of Scripts} 

In cybersecurity, semi-supervised learning has consistently shown promise for reducing manual labeling requirements. While significant work has focused on Portable Executable malware analysis~\cite{joyce2023motif, CybIN, Corlatescu-NeurIPS-2023, Anderson-ArXiv-2018}, script-based threats present unique challenges.
\citet{Alam-ArXiv-2025} used pseudo-labeling to retrain malware classifiers on their own predictions, while \citet{Feng-IEEE-2025} demonstrated the effectiveness of LLMs in Android malware detection through feature extraction and prompt engineering.
Regarding our work's main focus, natural language explanations of cyber threats, \citet{Fujii-ESORICS-2024} showed that LLMs can accurately explain malware functions, though they highlighted the significant challenge of limited high-quality training data. \citet{Lu-IEEE-2025} tackled this limitation by creating a novel dataset called MalS, by using LLM-generated draft summaries with light human refinement, and achieving strong results through fine-tuning.

Our work extends these efforts by introducing an iterative self-training approach for script analysis across five languages. Unlike previous work, focused on binary classification or requiring human refinement, our method generates detailed security explanations, demonstrating consistent quality improvements through multiple iterations with minimal human input.

\begin{figure*}[!t]
\centering
\includegraphics[width=0.98\textwidth]{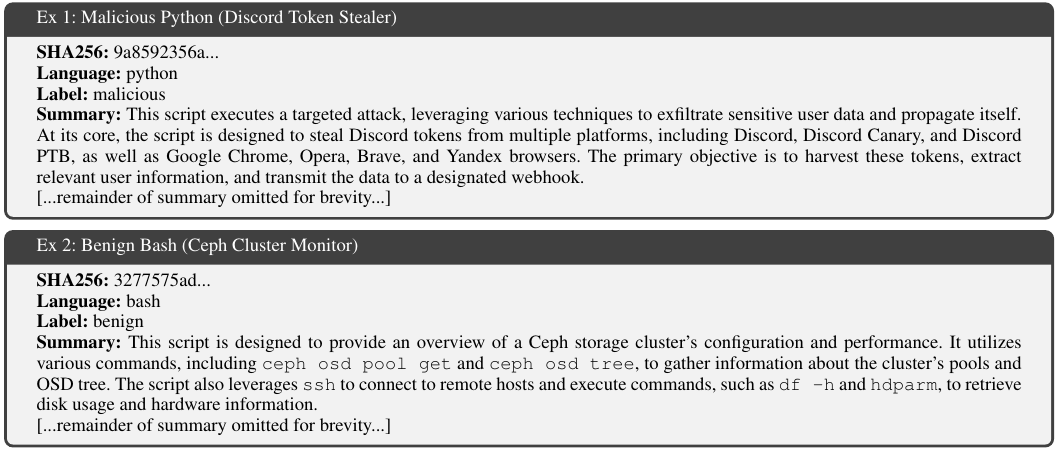}
\caption{Representative examples from our dataset showing malicious and benign scripts. Seed and test set summaries are generated using LLMs informed by SANDBOX detonation reports, while training set summaries use our iterative methodology without detonation data. Code references are formatted in \texttt{monospace} for clarity.}
\label{fig:example_seed_summaries}
\end{figure*}

\section{Dataset}
\label{sect:data}

We first describe our high-quality seed dataset used for initial model training. We then explain how we collected a larger unlabeled dataset from Virus Total, which is used in our iterative learning approach. Finally, we present the test set used for evaluation. 

\subsection{Seed Dataset}
\label{subsect:data_seed_gen}

Given the scarcity and cost of expert labeling, we followed  prior research~\cite{Li-ICLR-2024,Wang-ArXiv-2022} and composed a seed dataset of 900 scripts from Virus Total. This dataset supports three tasks: malware detection, scripting language classification and natural language description generation. Table~\ref{tab_data_distrib} shows the distribution across languages and maliciousness labels, while Figure~\ref{fig:example_seed_summaries} provides example entries with their corresponding descriptions.

To ensure quality of the malware tags, we combined multiple validation sources: (1) YARA rules for detecting known malicious patterns~\cite{Lockett-CoRR-2021}, (2) SANDBOX detonations for dynamic analysis~\cite{Guven-IJCESEN-2024} and (3) expert manual verification.
For generating high-quality descriptions of the scripts' behavior, we leveraged SANDBOX detonation reports for technical context and employ Llama-3.3-70B-Instruct (with the temperature $\tau\!=\!0.3$).
We note that the manual verification was conducted by subject matter experts (SMEs) with specialized cybersecurity expertise. They reviewed both the script content and corresponding SANDBOX detonation reports to ensure accurate classification of malicious and benign samples. This approach follows prior research methodologies~\cite{Li-ICLR-2024, Corlatescu-NeurIPS-2023,Wang-ArXiv-2022,Anderson-ArXiv-2018}, while addressing the challenges of obtaining expert-labeled data.

\subsection{Training Data Collection}
We collected 157,126 scripts from Virus Total, across five languages commonly associated with cybersecurity threats \cite{Feng-IEEE-2025a,Li-SIGSAC-2024,Srinivasan-STCR-2023,Sohan-IEEE-2020}: Bash (sh), Batch (bat), JavaScript (js), PowerShell (ps) and Python (py). The initial corpus distribution is shown in Table~\ref{tab_data_distrib} (first section). While the data appears balanced between benign and malicious samples, these labels -- derived from an internal YARA rules powered noisy labeling automation -- are shown for reference only and were not used for training due to their inherent noise.

Through multiple filtering and validation steps, including consistency checks and quality thresholds, we refined this initial corpus to 101,277 examples to get our final training dataset, as illustrated in the second section of Table~\ref{tab_data_distrib}. The complete filtering methodology and validation process are described in detail in Section~\ref{sect:method}.

\begin{figure}[!t]
\centering
\includegraphics[width=0.8\columnwidth]{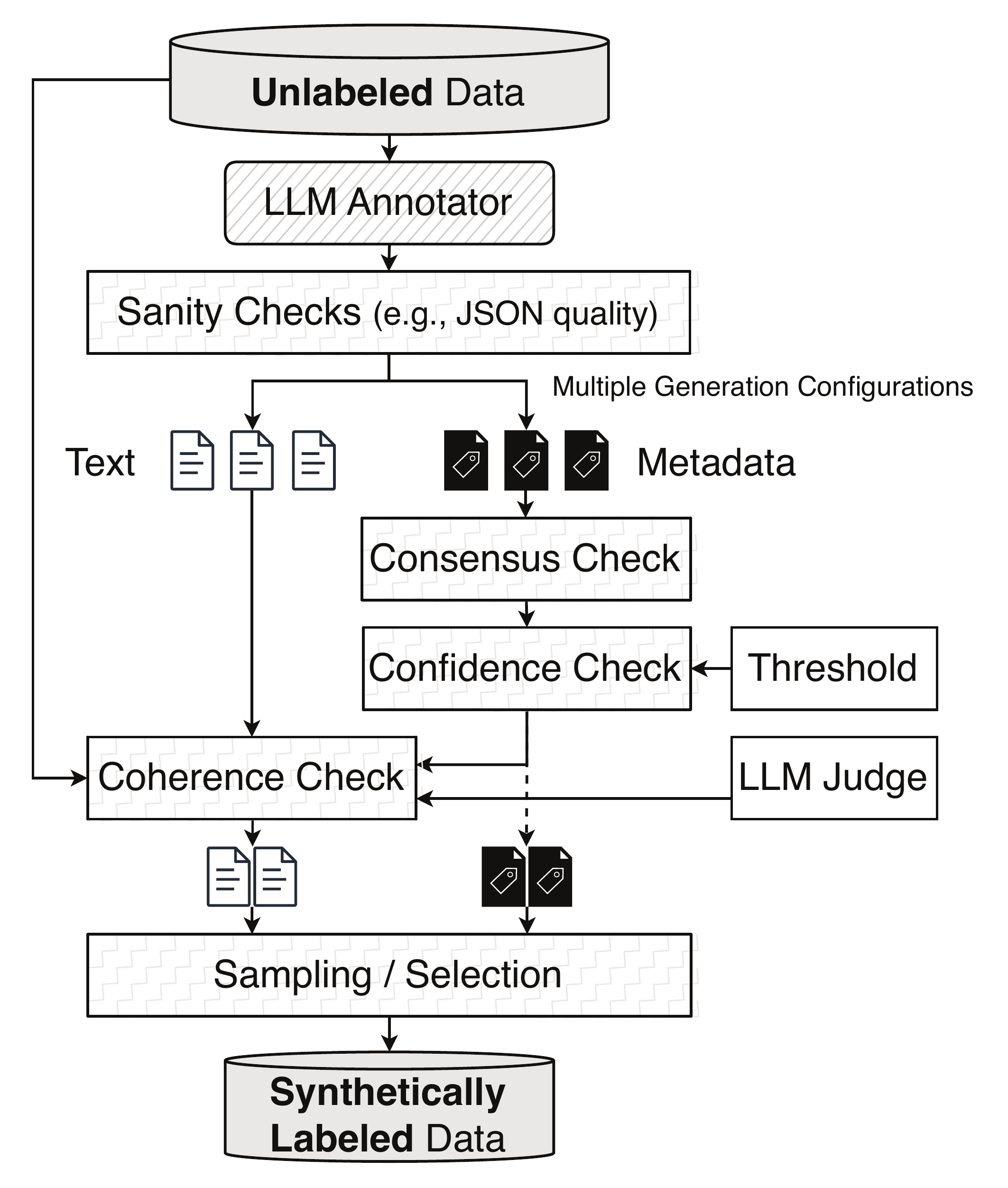}
\caption{Pseudo-label quality filtering pipeline. The LLM annotator generates metadata and natural language explanations using multiple configurations (e.g.~temperature settings) to ensure reliability. Samples pass filtering if metadata remains consistent across configurations and meets confidence thresholds. An LLM judge then verifies coherence between generated explanations and metadata labels.}
\label{fig:filtering}
\end{figure}

\subsection{Test Set}
\label{subsect:data_test}

Following benchmarking research recommendations~\cite{Guo-NeurIPS-2024}, we sample 700 additional samples per language, creating a test set of 3,636 samples. The quality of the test set assurance follows the methodology described in Subsection~\ref{subsect:data_seed_gen}, generating behavior-informed script summaries and validated scripting language and maliciousness labels. Analysis reveals that obfuscation techniques are present in at least 6\% of test samples, reflecting real-world malware.

Finally, we release the three subsets described through this section: a filtered training set of 101,277 samples, a high-quality seed set of 900 samples, and a test dataset of 3,636 samples. Each entry contains the script's SHA256 identifier, maliciousness and scripting language labels, as well as the generated script summaries. The seed and test sets feature SANDBOX-informed summaries, while the training set summaries follow our iterative methodology, as detailed in Section~\ref{subsect:method_iterative_data_extension}. To ensure responsible disclosure, we do not include script contents in our public release, though all samples remain accessible through VirusTotal's platform to authenticated users who have completed requisite background verification and compliance checks.


\section{Methodology}
\label{sect:method}

Our self-training methodology (Figure~\ref{fig:methodology}) leverages a small, high-quality seed dataset to fine-tune an LLM, which is then used to generate pseudo-labels for a larger corpus of unlabeled scripts. Through iterative refinement, we expand the training data in a self-supervised manner. This section details two key components: iterative dataset extension and self-training strategy. We evaluate our approach on three tasks: malware classification, programming language identification (collectively referred to as metadata in upcoming sections), and natural language behavior explanation.

\subsection{Iterative Dataset Extension}
\label{subsect:method_iterative_data_extension}

This subsection touches on our synthetic data generation process and subsequent dataset refinement (Figure~\ref{fig:filtering}). We begin by fine-tuning a large language model on our seed dataset to create LLM Annotator V1, which we use to generate pseudo-labels for our initial 157,126 unlabeled scripts. Given the limited training data, we implement rigorous filtering to ensure quality, as described below.

\subsubsection{Initial generation:} Using Annotator V1 (Llama-3.3-70B-Instruct fine-tuned on seed data), we generate explanations and metadata (maliciousness labels and language identification) in JSON format for 157,126 scripts across multiple temperature settings.

\subsubsection{Quality filtering:} 
\begin{figure}[t]
\centering
\includegraphics[width=0.99\columnwidth]{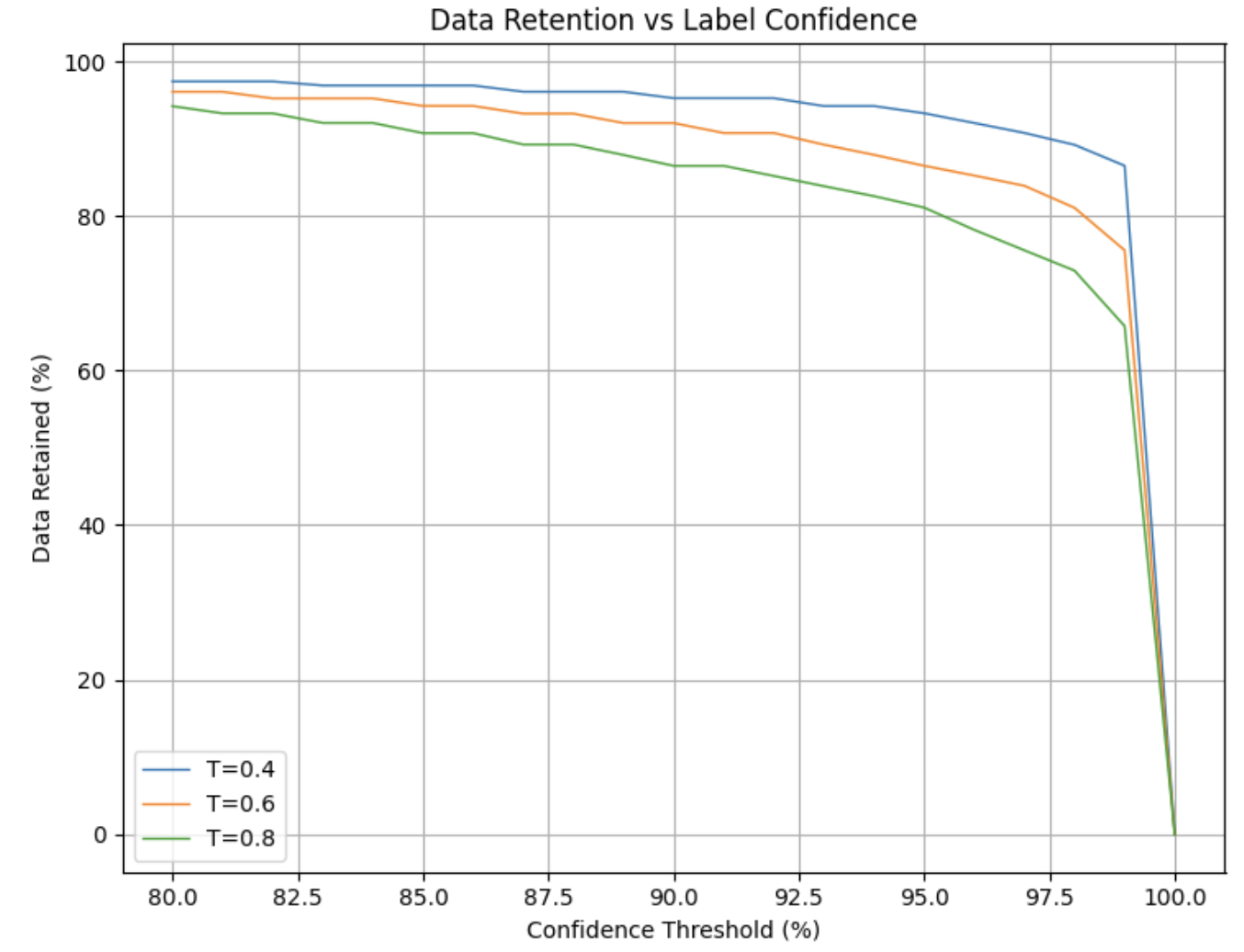}
\caption{Training data retention based on confidence scores. Distribution of model confidence scores, derived from logit probabilities of maliciousness predictions. In our case, a 90\% threshold balances quality and quantity, though this value is task-specific. Best viewed in color.}
\label{fig:label_retention_confidence}
\end{figure}

Our filtering pipeline consists of multiple stages. Initial cleanup removed empty responses (9,095), truncated summaries (17) and incomplete JSONs (7,825), reducing the dataset by 10.76\%. Consistency checks across three temperatures (0.4, 0.6, 0.8) eliminated samples with inconsistent malicious labels, reducing data by another 14.28\%. Using Phi-3.5-Mini, selected for its 99.5\% accuracy on our test set, we removed 219 samples (0.22\%) with discrepancies between summaries and maliciousness labels. Finally, we applied a 90\% confidence threshold on logit probabilities, with most values clustering near 1 (Figure~\ref{fig:label_retention_confidence}). The resulting filtered dataset contains 101,277 examples (Table~\ref{tab_data_distrib}). This filtering strategy balances correctness with dataset size, though threshold values may need adjustment for different applications.

\begin{table}[t]
\centering
\begin{tabular}{ccccccc}
    \multicolumn{7}{c}{Language Detection Accuracy (\%)} \\
    Model  &  sh &  bat &  js &  ps & py & Avg.\\
    \hline
    Base     & 89.2 & 92.3 & 98.4 & 89.9 & 100    & 93.7\\
    V1 & 98.0 & 94.2 & 99.5 & 92.7 & 99.8 & 96.8\\
    V2 & 99.3 & 94.9 & 99.8 & 91.8 & 100    & 97.1\\ 
    \\
    \multicolumn{7}{c}{Malware Detection Accuracy (\%)} \\
    Model  &  bash &  batch &  js &  ps & py & Avg.\\
    \hline
    Base     & 92.4 & 52.7 & 90.8 & 94.2 & 89.8 & 83.1\\
    V1 & 93.2 & 77.9 & 90.6 & 95.3 & 91.8 & 89.3\\
    V2 & 96.3 & 82.4 & 92.2 & 95.3 & 92.6 & 91.5\\ 
\end{tabular}
\caption{Accuracy percentages for language detection and malware detection across three model versions (Base: pretrained Llama-3.3-70B-Instruct, V1: first iteration, V2: second iteration) for five scripting languages. Note the significant improvement in batch script malware detection (52.7\% increase from Base to V2) and consistent gains across languages through our iterative training.}
\label{tab:detection_accuracy}
\end{table}

\subsection{Self-Training Strategy}
\label{subsect:self_training_strategy}

Our full self-training methodology is illustrated in Figure~\ref{fig:methodology}. To train the LLM annotators, we used an instruction-tuning approach to specialize these models for our tasks -- i.e.~malware detection, language classification and summary generation. Given the relatively small size of the datasets and large architectures used throughout our experiments, we decided to use Low-Rank Adaptation (LoRA)~\cite{Hu-ICLR-2022}, known to efficiently adapt large models with minimal trainable parameters. Each iteration follows the same training process but uses different datasets: Annotator V1 is trained on the seed data, while Annotator V2 incorporates both seed and V1-generated samples. To prevent overfitting, we monitor the validation loss on a held-out subset, training until convergence (11 epochs for V1, trained on the smaller seed dataset; 13 epochs for subsequent iterations).

\section{Experiments}
\label{sect:experiments}

In this section, we succinctly present our experimental setup and configuration parameters, followed by a detailed analysis of our self-improving method's performance and results.

\subsection{Experimental Setting}

Through our experiments, we fine-tuned Llama-3.3-70B-Instruct using Low-Rank Adaptation (LoRA) on NVIDIA H100 Mega GPUs. Base hyperparameters included: 16k context size, batch size of 1, learning rate of 0.0001, weight decay of 0.001, and warmup ratio of 0.05, with gradient checkpointing and packing enabled for memory optimization.
Annotator V1 used LoRA parameters (rank of 8, $\alpha\!=\!16$, dropout rate of 0.05) and target modules set to ``all-linear'', training for 15 epochs with epoch 11 providing optimal performance. Annotator V2 maintained the base configuration but doubled the LoRA rank and $\alpha$ (to 16 and 32, respectively), and increased dropout to 0.1. Of its 20 training epochs, epoch 13 proved optimal.

Best-performing checkpoints were selected based on validation loss convergence to prevent overfitting, then evaluated on our held-out test dataset. Additional details are provided in supplementary materials.

\subsection{Results}
\label{subsect:results}

We evaluate our annotator models, V1 and V2, through both qualitative and quantitative analysis. For detection capabilities (i.e.~language identification and malware detection), we use our test set of approximately 3,600 samples (Subsection~\ref{subsect:data_test}), comparing performance against the base Llama-3.3-70B-Instruct model. For script comprehension, we assess 215 samples under 3KB across the five languages in scope (Bash, Batch, Javascript, PowerShell, and Python). Both human and LLM judges performed pairwise evaluations of generated summaries, with the option to rate V1 and V2 outputs as equivalent when appropriate. 

\subsubsection{Detection capabilities:}

\begin{table}[t]
    \centering
    \begin{tabular}{lccc}
        \multicolumn{3}{c}{Model Evaluation Results (V1 vs.~V2)} \\
        Evaluator & V1 & V2 \\
        \hline
        Human Annotators & 46.26\% & 44.85\% \\
        Llama-3.3-70B-Instruct & 48.85\% & 51.15\% \\ 
        Claude 3.7 & 47.57\% & 52.43\%\\
        Phi-3.5-Mini-Instruct & 56.22\% & 43.78\% \\
        Mixtral-7x22B-Large & 52.07\% & 47.93\% \\
        GPT-4o & 51.15\% & 48.85\% \\
        GPT-5 (low) & 47.47\% & 52.53\% \\
    
    \end{tabular}
    \caption{Comparison of V1 vs.~V2 performance across different evaluators. Both human and LLM judges evaluated approximately 215 samples each. Percentages show the winrate of the model version where the samples were preferred. The table excludes 19 cases rated equivalent by human evaluators, as LLM judges did not have this option.}
    \label{tab:qualitative_assessments_comparison}
\end{table}

The base model, Llama-3.3-70B-Instruct, demonstrates strong language detection capabilities even prior to fine-tuning, achieving an average accuracy of 93.70\% across all scripting languages. It performs particularly well for Javascript and Python, but it is less robust for Bash (89.22\%). Its malware detection varies significantly, from over 90\% accuracy for Bash, Javascript and PowerShell, to 52.68\% for Batch scripts, averaging at 83.09\%.

Our fine-tuning process yielded substantial improvements in both detection tasks. V1 increased Bash language detection by 8.8\%, while V2 achieved over 99\% language accuracy for Bash and Javascript, and perfect 100\% for Python. Similarly, malware detection improved across all languages, with enhancements ranging from 1.1\% to 3.8\% for most languages, and a significant 29.9\% improvement for Batch.

McNemar's test~\citep{McNemar-Psychometrika-1947} confirmed that the improvements of V2 over V1 are statistically significant ($p<10^{-5}$), with V2 correcting 110 false positives and 49 false negatives from V1's predictions.

These results highlight the effectiveness of our fine-tuning approach in enhancing the model's ability to correctly identify both scripting languages and malicious samples, particularly addressing specific weaknesses in the base model's detection capabilities for previously challenging script types.

\begin{figure*}[t]
\centering
\includegraphics[width=0.98\textwidth]{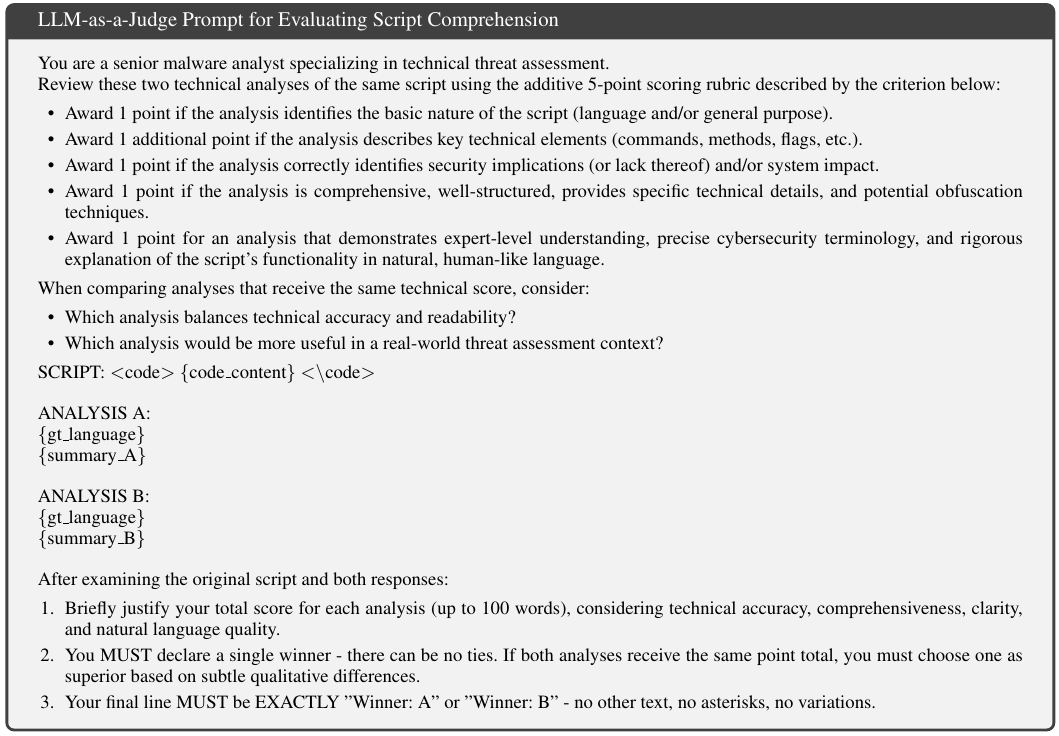}
\caption{Full evaluation prompt used in evaluating Script Comprehension capabilities for model V1 and V2. This same prompt was used in all four judge models: Claude 3.7 Sonnet, Llama 3.3 70B Instruct, Phi 3.5 Mini Instruct, and Mixtral 7x22B Large.}
\label{fig:llm_judge_prompt}
\end{figure*}

\subsubsection{Human evaluation of script comprehension:}

We evaluated the generated summaries through both human and LLM judges, with human evaluations conducted blindly to prevent bias. Each human judge assessed approximately 20 randomized pairs of summaries, providing their preference and reasoning behind the selection. 

Human evaluators found that both models produced similar and factually accurate summaries in 50.42\% of the cases, demonstrating comparable script comprehension capabilities. Preferences between models primarily reflected stylistic differences (e.g.~verbosity, level of detail, information flow and structure), rather than factual accuracy. Despite these personal preferences influencing decisions, the manual evaluation showed a marginal preference for V1 at 50.46\%, with V1 being selected 54 times compared to V2's 53 selections, and 9 instances where both models were rated equally effective. On a less positive note, we observed 27 total instances of hallucinations across both models (V1: 13, V2: 14), with varying distribution across programming languages. V1 had higher susceptibility to hallucination when processing PowerShell (5 instances) and Batch/Python scripts (3 each), with perfect reliability for Javascript samples. In contrast, V2 struggled primarily with Batch scripts (8 instances), but improved on Python (only 1 instance). 

\subsubsection{LLMs judging script comprehension:}

To complement human evaluations, we employed multiple LLM judges. Using a standardized prompt (Figure~\ref{fig:llm_judge_prompt}), each LLM evaluated summaries for script identification, technical coverage, security impact, composition quality and cybersecurity terminology accuracy. As indicated in Table~\ref{tab:qualitative_assessments_comparison}, three of the six LLM judges aligned with human evaluators in preferring V1, Claude 3.7 Sonnet and GPT-5 notably diverged, showing a stronger preference for V2. The other LLM judges consistently favored V1. The split in LLM preferences suggests that different evaluation frameworks may emphasize different qualities in script summaries, when factual accuracy is consistently high. Notably, Phi-3.5-Mini-Instruct, with its smaller architecture (3.8B parameters), was unable to complete the full evaluation set, likely due to its limited capacity compared to larger models.

\section{Discussion}
\label{sect:discussion}

This section extends beyond the evaluation presented in Subsection~\ref{subsect:results}, focusing on two key aspects. First, we discuss how SANDBOX-informed seed summaries improve generation quality through pattern preservation across iterations. Second, we present detailed human observations on stylistic evolution between Annotator V1 and V2, providing deeper insights into our models' generation characteristics.

\subsubsection{Knowledge transfer from detonation reports:} To assess the effectiveness of SANDBOX-informed seed annotations, we analyzed behavioral pattern preservation between seed-summaries and descriptions generated by Annotator V1. Table~\ref{tab:traces_of_detonations} presents a few examples of shorter common patterns found in both datasets, with additional longer examples including \textit{``Queries sensitive IE security settings''}, \textit{``Calls an API typically used to create an HTTP or FTP session''} and \textit{``The input file contains long base64 encoded strings''}. Concretely, over $50$ seed summaries, and over $1,300$ V1 summaries contained exact matches. The presence of these patterns in over 50 seed summaries and 1,300 V1-generated descriptions demonstrates successful transfer of SANDBOX-derived knowledge through fine-tuning. 

Especially, for annotating large datasets, our approach offers significant advantages over traditional detonation analysis. While SANDBOX detonations are expensive, time-consuming and dependent on specific environmental conditions, LLM-based analysis provides faster, and, depending on the use-case, even more cost-effective insights. Furthermore, static analysis can identify potential behaviors even when dynamic execution fails, particularly for scripts dependent on inaccessible resources, sandbox-aware malware, or specific environmental configurations.

\begin{table}[t]
\centering
\begin{tabular}{lccc}
    Phrase & Seed & V1 \\
    \hline
    \textit{Writes data to a remote process} & 10 & 141 \\
    \textit{Writes registry keys} & 9 & 32\\
    \textit{Creates new processes} & 8 & 98 \\
    \textit{Modifies proxy settings} & 8 & 48 \\
    \textit{Deletes registry keys} & 8 & 26 \\
    \textit{Executes a shell command} & 7 & 214 \\
    \textit{Executes a JavaScript file} & 5 & 269 \\
    \textit{Queries process information} & 5 & 24 \\
    \textit{Queries sensitive IE security settings} & 5 & 30 \\
    \textit{Drops cabinet archive files} & 5 & 9 \\
    \textit{Invokes the C\# compiler} & 3 & 10 \\
    \textit{Searches for file content} & 2 & 256 \\

\end{tabular}
\caption{Examples of behavioral patterns present in both SANDBOX-informed seed summaries and subsequent V1-generated descriptions.}
\label{tab:traces_of_detonations}
\end{table}

\subsubsection{Summary generation style progression:}
Our human evaluation revealed distinct stylistic patterns between V1 and V2 summaries despite their similar technical accuracy. V1 maintained a more technical and conservative approach, focusing on specific implementation details like command flags, memory allocation parameters, and API calls. V2 produced more readable summaries with improved narrative flow, while showing several noteworthy characteristics.
First, V2 demonstrated a tendency toward dramatic characterization, occasionally describing scripts as ``formidable threats'' and making broader assumptions about script intent. This was particularly evident in its treatment of benign scripts, where it sometimes overestimated security risks, while V1 maintained a more neutral stance. Annotators consistently noted this difference, often preferring V1's more measured approach for clean scripts. Second, while V2's summaries were generally more readable and fluid, they occasionally exhibited redundancy, particularly when describing obfuscation techniques or encoding schemes. V1's concise, technical descriptions were often preferred despite being less narrative in style. These observations suggest a trade-off between technical precision and natural language fluency across iterations. While V2 improved readability and narrative flow, it also introduced a bias toward assuming malicious intent and occasionally sacrificed precision for expressiveness. This pattern was consistent across multiple annotators, who frequently noted that their preferences were driven by summary style rather than technical accuracy, as both models maintained similar levels of factual correctness. Importantly, this V1/V2 trade-off serves distinct operational needs: V1 offers technical precision for engineers and malware analysts requiring detailed implementation specifics, while V2 prioritizes readability for broader audiences, without sacrificing classification performance.

\section{Limitations}
\label{sect:limitations}
Our approach faces three limitations. First, model context length constraints prevent analysis of longer scripts, particularly affecting PowerShell and JavaScript samples with extensive code blocks. Second, language detection accuracy (97.14\% in V2) shows minimal improvement across iterations, suggesting a performance ceiling. Third, we observe a clear trade-off between readability and technical precision, with V2 favoring natural language flow at the occasional expense of technical accuracy, a difference that human evaluators found subtle and largely stylistic.

\section{Conclusions}
\label{sect:conclusions}

In this work, we demonstrated how self-improving language models can enhance malware script analysis, achieving statistically significant improvements in classification and summary generation with minimal human supervision. Our approach notably improved batch script malware detection from near-random (52.7\%) to a robust (82.4\%) accuracy across two iterations. To support reproducibility and future research, we release a dataset of more than 100K examples, spanning five scripting languages with balanced representation of benign and malicious samples. Furthermore, we make available an evaluation framework, combining quantitative metrics with human and LLM-based qualitative assessment, establishing a robust benchmark for future work in this domain. Results confirm that LLMs can effectively learn to identify malicious behavior patterns from limited seed data, substantially reducing reliance on costly SANDBOX detonations and expert annotations.
 
\section*{Ethical Statement}
To balance research accessibility with security concerns and responsible disclosure practices, we release only SHA256 identifiers of scripts along with the associated artefacts derived from our research, requiring VirusTotal authentication for accessing the scripts content. Thus, aligned with established security research practices and industry standards, our publicly available dataset includes comprehensive metadata, labels and summaries, enabling research without exposing harmful code, while maintaining reproducibility.

\section*{Acknowledgments} The authors would like to thank Bilal Issa, Ioana Angela Ileni, Mihai Alestar and Monica Pascu for their input and expertise. The authors would further like to thank Sven Krasser, Cheryl Houser, Ciprian Bejan, Dragos Corlatescu, Alexandru Dinu, John Zuehlke and Marian Radu for their help and support.

\bibliography{aaai2026}

\appendix
\section*{Appendix}

\subsection*{Related Work}

Table~\ref{tab:datasets_comparisons} showcases several datasets that have been proposed for malware analysis in scripts. However, most of these data sets focus on single programming languages and lack natural language explanations. In the continuation of this subsection, we categorize the existing work based on the primary scripting languages they cover.

\paragraph{PowerShell-focused datasets:} 
The largest PowerShell-specific datasets include the works of \citet{Fang-Neurocomputing-2021} and \citet{Hung-JIT-2024}, with approximately 4,200 and 4,100 malicious samples, respectively, both having comparable numbers of benign samples. \citet{Varlioglu-IEEE-2024} introduced a smaller dataset of 200 malicious PowerShell scripts, though without benign counterparts.

\paragraph{Python-centric collections:}
Python script datasets largely focus on package-level analysis. \citet{Guo-ASE-2023} presented the largest Python-specific collection with 4,669 malicious and 549 benign samples. Other notable contributions are those of \citet{Samaana-SIGAPP-2025} and \citet{Vu-ArXiv-2022}, who focused on PyPI packages, containing 138 and 168 malicious samples, respectively, complemented by larger benign sample sets.

\paragraph{Multi-language datasets:}
Few datasets cover multiple scripting languages. \citet{Ladisa-ACSAC-2023} analyzed both Python and JavaScript scripts, while \citet{Erdemir-ArXiv-2024} provided the largest multi-language dataset, covering Python, Bash, and Perl with 48,259 malicious and 363,998 benign samples.

\paragraph{Limitations of existing datasets:}
Existing datasets exhibit several significant limitations. Most notably, they typically focus on a single scripting language, limiting their applicability in real-world scenarios where multiple languages are commonly used together. Furthermore, none of the existing datasets include Batch scripts, creating a notable gap in Windows-based script analysis. The absence of natural language explanations across all existing datasets makes it challenging for researchers to understand and validate the malicious behaviors being studied. Additionally, many of these datasets suffer from significant class imbalance between malicious and benign samples, potentially impacting the effectiveness of machine learning models trained on this data.

Our work addresses these limitations by introducing a comprehensive dataset covering five scripting languages (PowerShell, Bash, Batch, JavaScript, and Python) with 2,368 malicious and 2,168 benign samples, maintaining better class balance. Uniquely, our dataset includes natural language explanations for all samples, facilitating better understanding and analysis of malicious behaviors.

\begin{table*}[tbp]
    \centering
    \resizebox{.85\width}{!}{
    \begin{tabular}{l|c|c|l|c}
        \textbf{Dataset} & \textbf{Malicious samples} & \textbf{Benign samples} & \textbf{Languages Covered} & \textbf{Explanations} \\
        \hline
        \cite{Fang-Neurocomputing-2021} & 4,202 & 4,316 & PowerShell & No \\
        \cite{Hung-JIT-2024} & 4,100 & 5,189 & PowerShell & No \\
        \cite{Varlioglu-IEEE-2024} & 200 & 0 & PowerShell & No \\
        \cite{Samaana-SIGAPP-2025} & 138*  & 5,193*  & Python & No \\
        \cite{Ladisa-ACSAC-2023} & 194* & 1640* & Python, JavaScript & No \\
        \cite{Guo-ASE-2023} & 4,669* & 549* & Python & No \\
        \cite{Vu-ArXiv-2022} & 168* & 2,416 & Python & No\\
        \cite{Fang-Wiley-2021}& 515* & 1,511* & Python & No\\
        \cite{Erdemir-ArXiv-2024} & 48,259 & 363,998 & Python, Bash, Perl & No \\
        \hline
        Ours (seed + eval) & 2368 & 2168 & PowerShell, Bash, Batch, JavaScript, Python & Yes \\
    \end{tabular}
    }
    \caption{Comparison of publicly available script datasets with our work (AutoMalDesc). Entries marked with * represent package-level scripts (e.g.~PyPI packages for Python). Existing datasets predominantly focus on single scripting languages and lack both Batch script coverage and natural language explanations. The table shows sample counts for malicious and benign scripts across different programming languages.}
    \label{tab:datasets_comparisons}
\end{table*}

\subsection*{Seed Data Generation}

This subsection details our systematic approach to prompt engineering for generating high-quality script explanations in the seed dataset. Through multiple iterations of prompt development and testing, we identified and addressed several significant challenges in achieving consistent and accurate malware assessments.

\paragraph*{Initial approach:}
We began with a minimalist prompt with 5-6 bullet-point instructions for analysis requirements and basic formatting guidelines. Experimental evaluations across multiple temperature settings (0.2, 0.3, 0.5) exposed two key limitations in the model's responses. First, the generated analyses exhibited mechanical repetition, with the model addressing each instruction point in a rigid, formulaic manner that lacked the natural flow characteristic of expert malware analysis. Second, the outputs were frequently undermined by excessive meta-commentary, where the model would interrupt its analysis with self-referential phrases such as \textit{"I will now summarize the script"} or \textit{"I must write only 1-2 paragraphs."} This behavior demonstrated the base model's inherent tendency to maintain its assistant persona rather than fully embodying the role of a malware analyst providing direct technical assessment.

\paragraph*{Paragraph-based approach:}
To address the mechanical responses and meta-commentary issues identified in our initial approach, we restructured the prompt into a more natural paragraph format. This revision organized the instructions into two distinct sections: \textit{``Content Instructions''} and \textit{``Prohibitions''} moving away from the bullet-point structure that had produced rigid responses. Our hypothesis was that presenting requirements in flowing prose, coupled with explicit instructions for fluid writing, would encourage the model to generate more naturalistic analytical responses. Contrary to our expectations, this modification resulted in degraded output quality. The model produced extremely condensed, list-like analyses that exhibited substantially reduced technical depth compared to our initial bullet-point approach. This unexpected regression in analytical quality suggested that the relationship between prompt structure and response sophistication was more complex than initially assumed, and that merely reformatting instructions into paragraphs was insufficient to achieve the desired expert-level analysis.

\paragraph*{Template-based approach:}
Building upon lessons learned from both the initial and paragraph-based approaches, our third iteration returned to a structured format, while introducing explicit analytical sections (e.g.~Technical Function, Security Impact) to guide the organization of the model's response. Testing with a temperature setting of 0.3 revealed new challenges that differed from our previous attempts. The model began incorporating section headers verbatim and increased its meta-commentary (e.g.~\textit{``The best answer is...''}), while also disregarding explicit prohibitions by including code snippets in its analyses. Our attempt to address these issues through stronger prohibitive instructions proved counterproductive, as the model became overly focused on avoiding forbidden behaviors rather than delivering substantive analysis. This overcorrection manifested in responses containing self-referential statements like \textit{``NO code snippets''} and \textit{``I mustn't include meta-commentary''}, alongside fragmented analyses that either terminated prematurely or contained redundant content. A significant improvement came from optimizing the input features by removing overrepresented Tactics, Techniques, and Procedures (TTPs) that occurred 20 times more frequently than other signatures,  generating excessive noise in the outputs. While this adjustment enhanced response quality, approximately 20\% of the generated analyses still exhibited quality issues, indicating the need for further prompt engineering refinements.

\begin{figure*}[!t]
\centering
\includegraphics[width=0.98\textwidth]{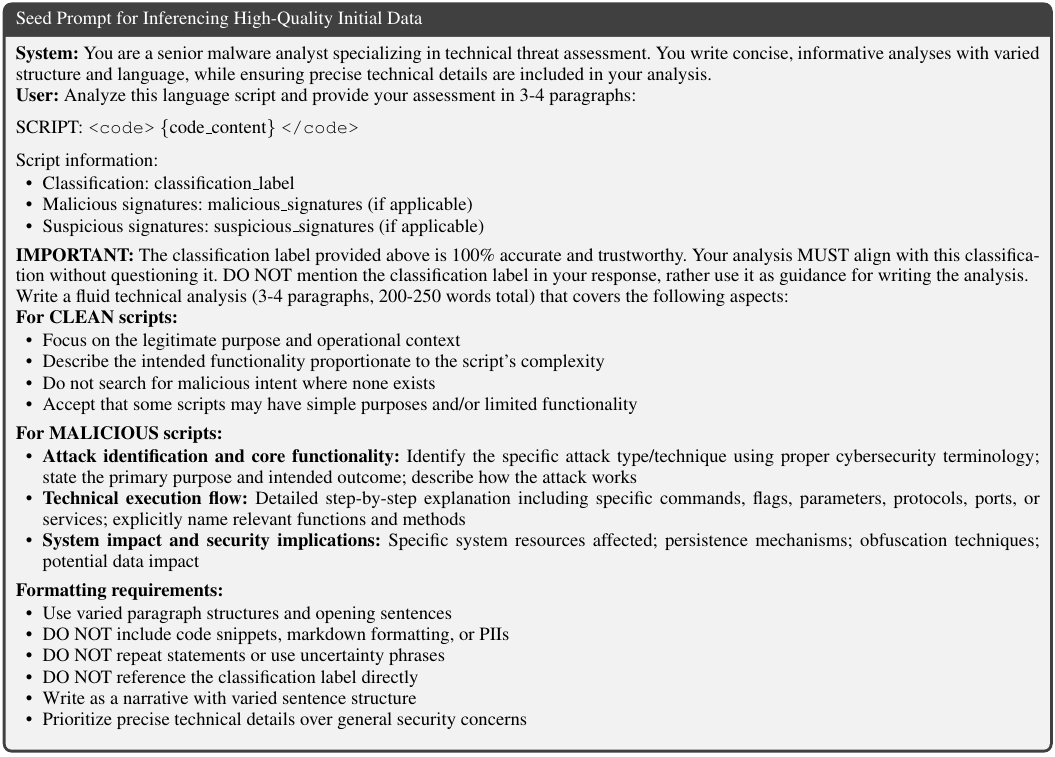}
\caption{Comprehensive seed prompt engineering for generating high-quality script analyses using LLama3.3-70B-Instruct. The prompt establishes a role-playing scenario where the LLM acts as a senior malware analyst, ensuring technical accuracy and professional analysis. It includes specific instructions for handling both clean and malicious scripts, with detailed requirements for analysis depth, structure, and content. The prompt leverages detonation report data by incorporating malicious and suspicious signatures as contextual input. Key components include: (1) system context setting for consistent analyst persona, (2) structured input format for script content and classification metadata, (3) explicit classification-dependent analysis requirements, (4) detailed formatting and style guidelines to ensure consistent, high-quality outputs, and (5) specific constraints to prevent common LLM analysis pitfalls such as uncertainty statements or direct classification references. This carefully crafted prompt structure ensures generated analyses maintain technical precision while providing comprehensive coverage of script functionality or malicious behaviors.}
\label{fig:seed_prompt}
\end{figure*}

\paragraph*{System prompt integration:}
After encountering persistent issues with response quality, meta-commentary and structural consistency across our first three approaches, we achieved a significant breakthrough by fundamentally reconsidering the model's operational context. Our fourth iteration introduced a system prompt that explicitly defined the AI's role as a \textit{``senior malware analyst specializing in technical threat assessment''}. This approach combined a detailed persona definition with a structured three-section analytical template and comprehensive formatting guidelines, addressing many of the mechanical and meta-commentary issues observed in previous iterations. While this modification substantially improved response clarity and reduced the meta-commentary that hindered earlier attempts, it revealed a critical accuracy problem: approximately 25\% of clean samples were mischaracterized as malicious, indicating serious hallucination issues where the model actively sought malicious indicators in benign scripts. Our attempts to mitigate this through temperature adjustments (0.7 and 0.9) were aimed to encourage more natural language variation, but these modifications had the opposite effect of reducing technical precision, while exacerbating the hallucination problem. This experience highlighted the delicate balance between maintaining technical accuracy and achieving natural language flow in automated malware analysis.

\paragraph*{Final optimized prompt:}
Building upon the insights gained from our previous iterations, particularly the hallucination issues identified in the system prompt approach, our final prompt engineering solution (detailed in Figure \ref{fig:seed_prompt}) implemented several crucial structural modifications. We eliminated the rigid numbered section format that had contributed to mechanical responses in earlier versions, instead adopting more fluid, natural language instructions. To address the persistent misclassification of benign scripts observed in our fourth iteration, we introduced specific content guidance for clean script analysis, explicitly directing the model to focus on legitimate functionality and avoid searching for non-existent malicious behaviors. The hallucination problem was further mitigated by incorporating authoritative statements about classification label accuracy, declaring them \textit{``100\% trustworthy and authoritative''} to prevent the model from second-guessing provided classifications. Additionally, we introduced explicit requirements for varied paragraph structures and diverse opening sentences to combat the repetitive patterns observed across all previous attempts. This comprehensive refinement of our prompt engineering approach resulted in significantly improved analysis quality, successfully balancing technical precision with natural language flow while maintaining accurate classification alignment. The final prompt structure effectively addressed the major challenges encountered throughout our iterative development process: mechanical responses, meta-commentary, hallucination, and repetitive patterns.

\subsection*{Dataset Characteristics and Token Distribution Analysis}

To provide comprehensive insight into the training data used for both LLM annotators and address potential concerns regarding context length limitations, we conducted exploratory data analysis on script token distributions across our datasets.

\begin{figure*}[tbp]
\centering
\includegraphics[width=0.85\textwidth]{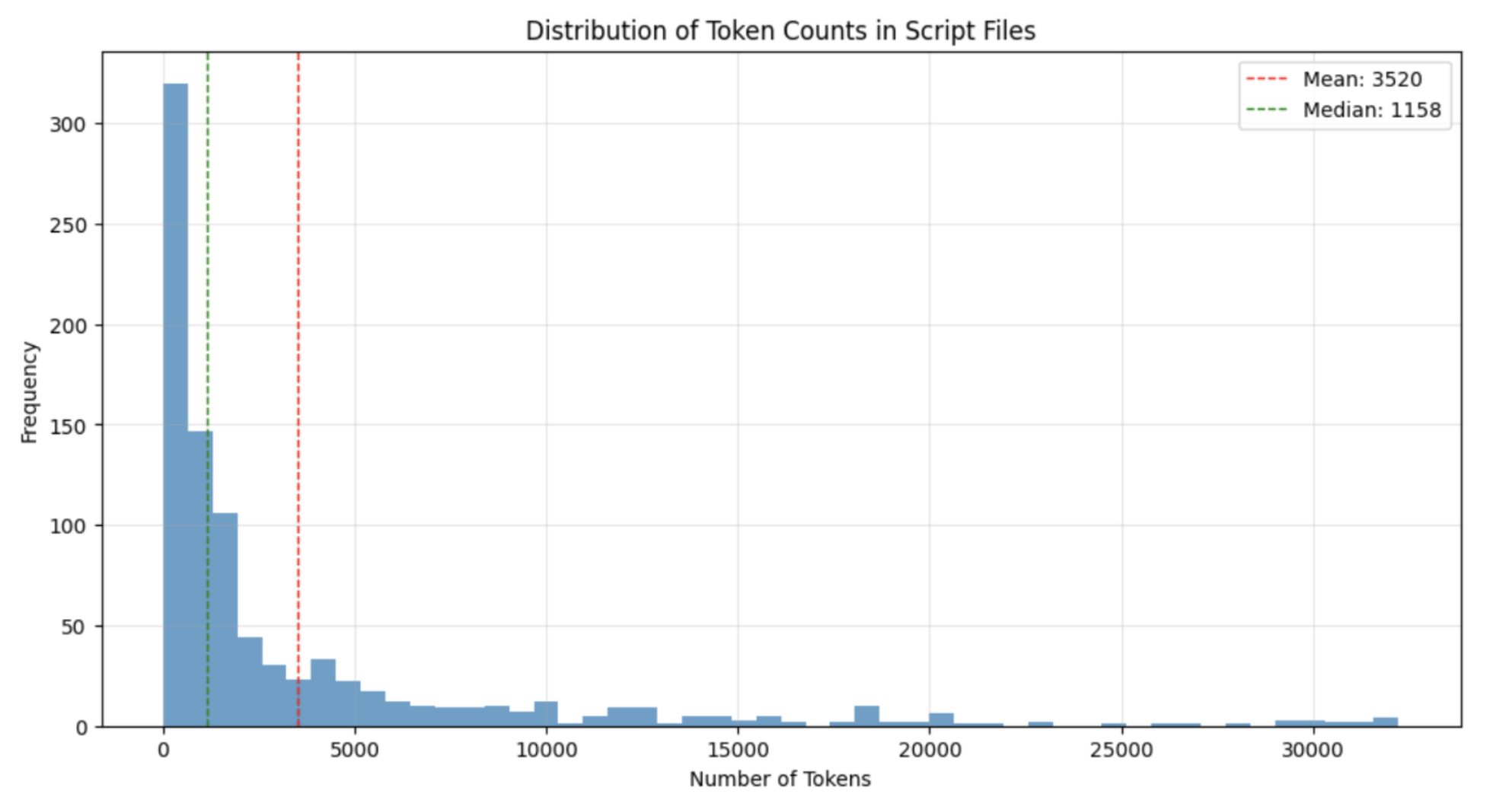}
\caption{Token count distribution for seed dataset (n=900).}
\label{fig:token_count_seed}
\end{figure*}

Figure \ref{fig:token_count_seed} presents the token count distribution for the seed dataset (900 samples). The majority of samples fall well within the 128k token context limit of our base model, with a mean of 3,520 tokens and a median of 1,158 tokens. This right-skewed distribution indicates that while most scripts are relatively compact, a small proportion of samples contain substantially more tokens.

\begin{figure*}[tbp]
\centering
\includegraphics[width=0.85\textwidth]{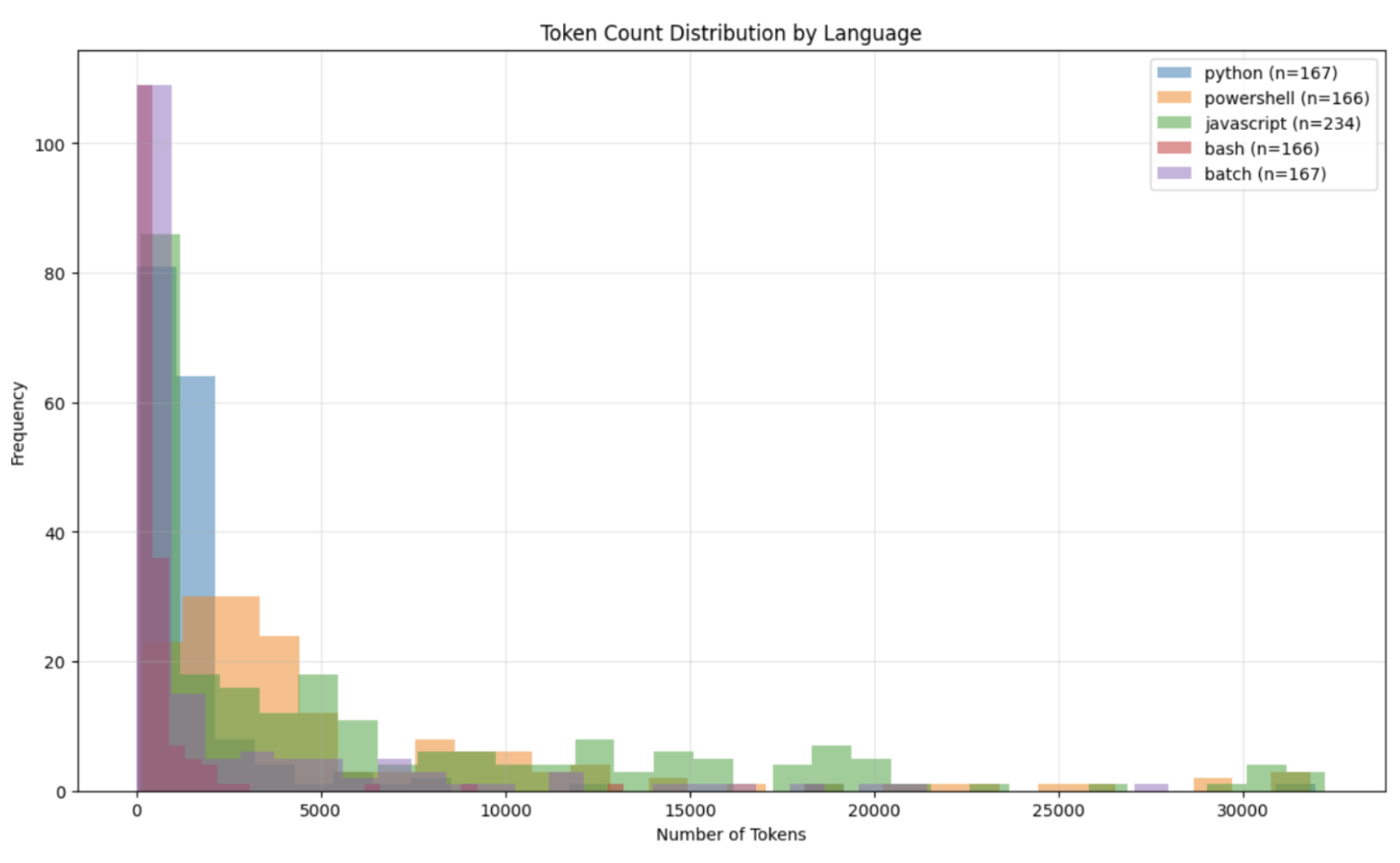}
\caption{Token count distribution by programming language for the seed dataset.}
\label{fig:token_count_seed_languages}
\end{figure*}

Breaking down the seed data by programming language (Figure \ref{fig:token_count_seed_languages}), we observe notable variation in script sizes across languages. JavaScript samples exhibit the longest scripts, with a few extending beyond 30,000 tokens, while PowerShell also contains a subset of lengthier samples. In contrast, Bash, Batch, and Python scripts tend to be more compact, with the majority clustering in the lower token ranges.

\begin{figure*}[tbp]
\centering
\includegraphics[width=0.85\textwidth]{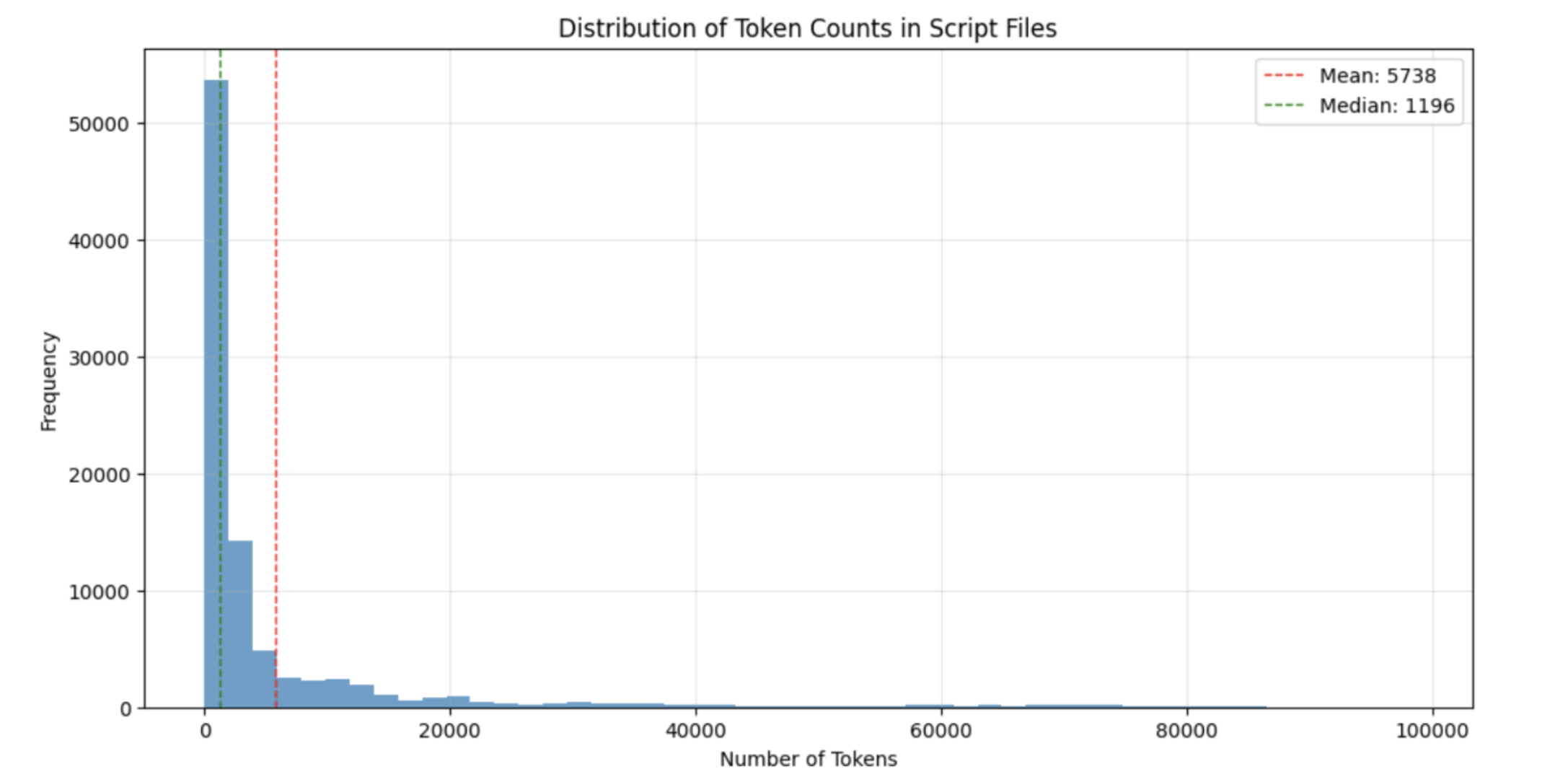}
\caption{Token count distribution for V2 training dataset (n=101K).}
\label{fig:token_count_v2}
\end{figure*}

\begin{figure*}[tbp]
\centering
\includegraphics[width=0.85\textwidth]{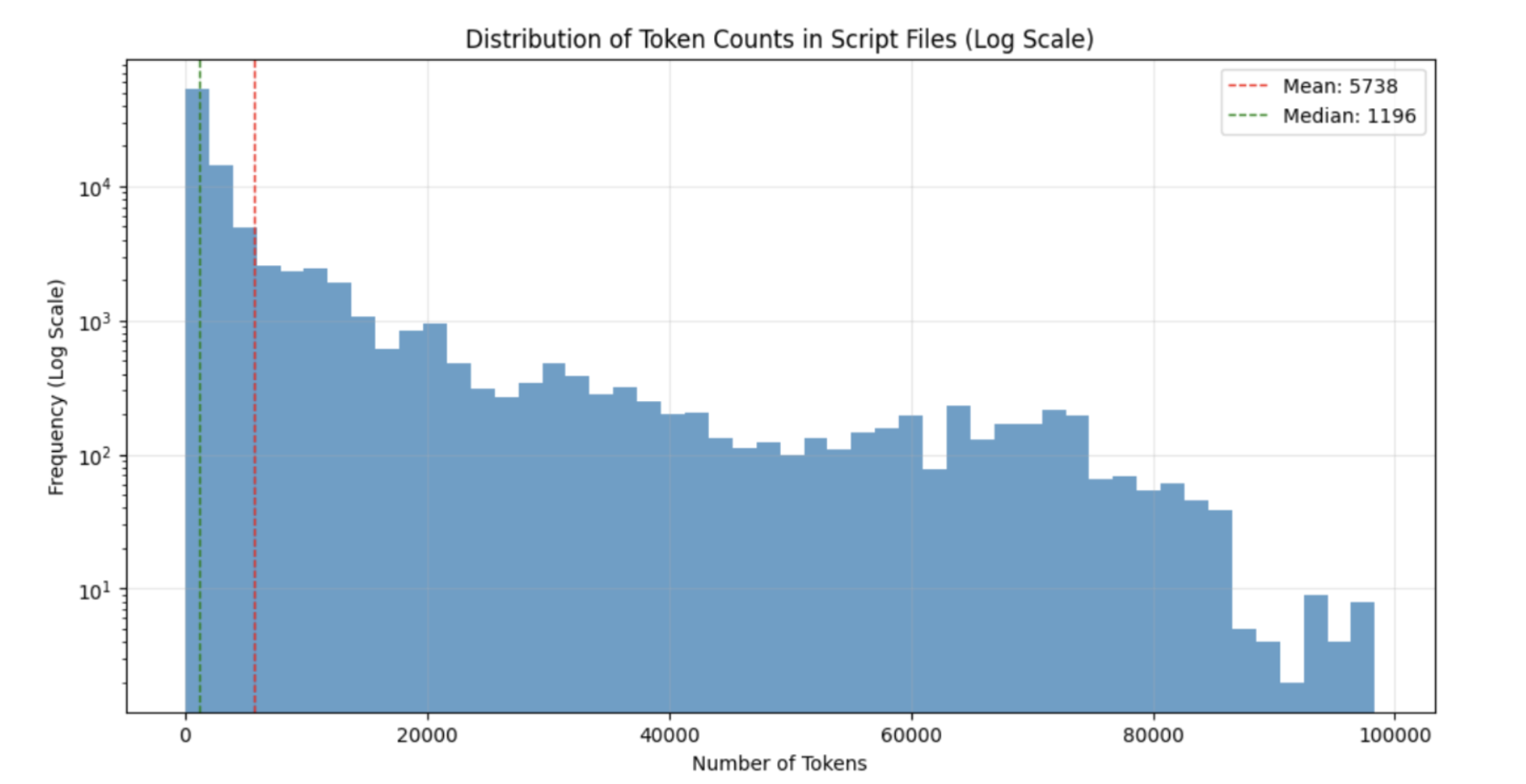}
\caption{Token count distribution for V2 training dataset displayed on logarithmic scale.}
\label{fig:token_count_v2_logscale}
\end{figure*}

For the substantially larger V2 training dataset, Figure \ref{fig:token_count_v2} illustrates the size distribution across approximately 101,000 samples. The distribution maintains a similar pattern, with a mean of 5,738 tokens and a median of 1,196 tokens. To better visualize and zoom in on the distribution, Figure \ref{fig:token_count_v2_logscale} presents the same data on a logarithmic scale, revealing that script frequency decreases exponentially with token count. This confirms that extremely long scripts represent a small fraction of the training data, mitigating concerns about context length bottlenecks.

\begin{figure*}[tbp]
\centering
\includegraphics[width=0.85\textwidth]{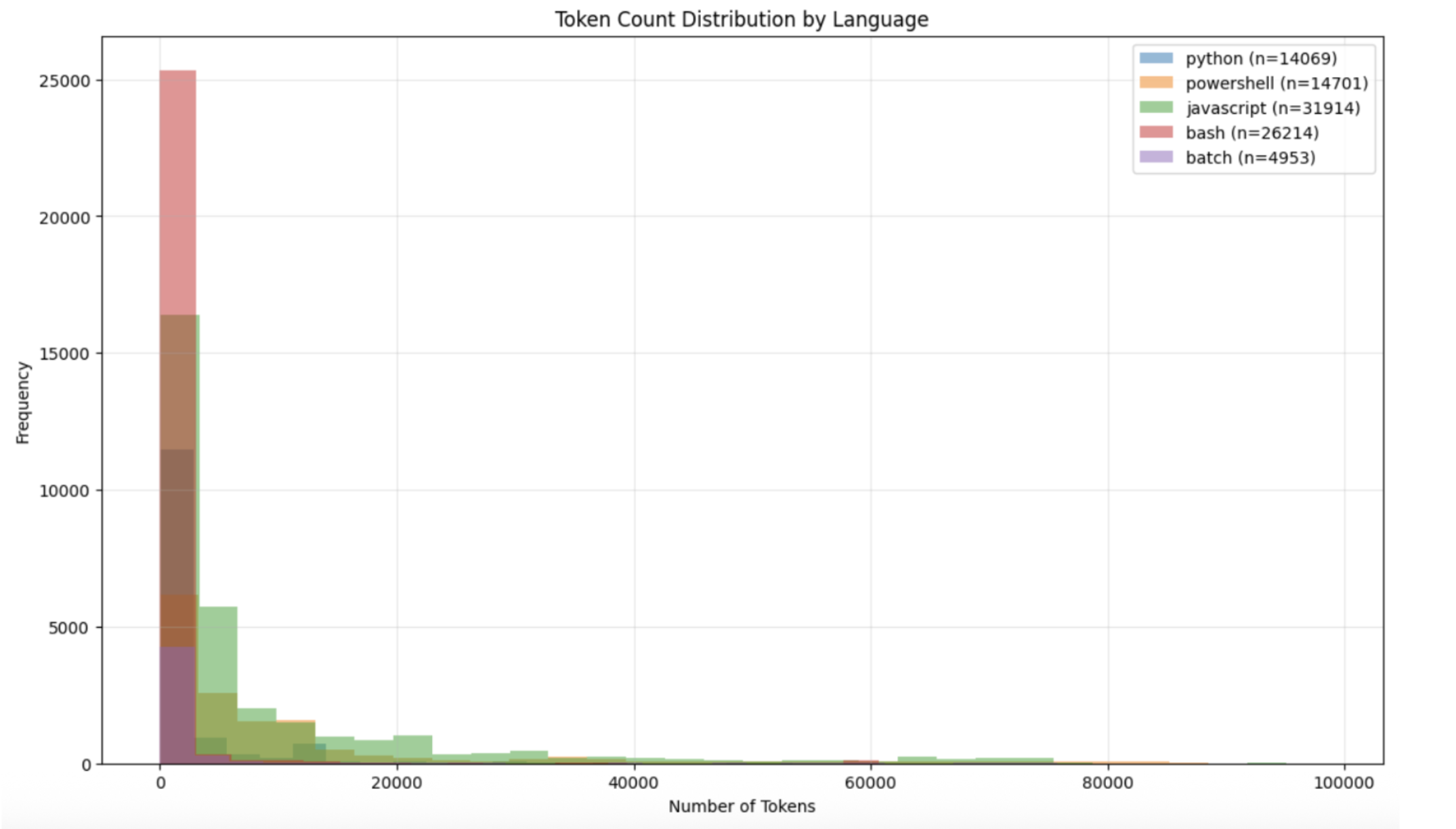}
\caption{Token count distribution by programming language for V2 training dataset.}
\label{fig:token_count_v2_languages}
\end{figure*}

The language-specific patterns observed in the seed data persist in the larger V2 training set (Figure \ref{fig:token_count_v2_languages}). JavaScript continues to dominate the upper range of token counts, with several samples approaching 80,000 tokens.

\subsection*{Data Filtering Insights}

While the previous subsection detailed our iterative prompt engineering process for generating seed and test datasets, this subsection presents our comprehensive data filtering methodology for explanations generated by the LLM Annotator V1. 

\subsubsection{Probability filtering:}

\begin{figure*}[!t]
\centering
\includegraphics[width=0.7\textwidth]{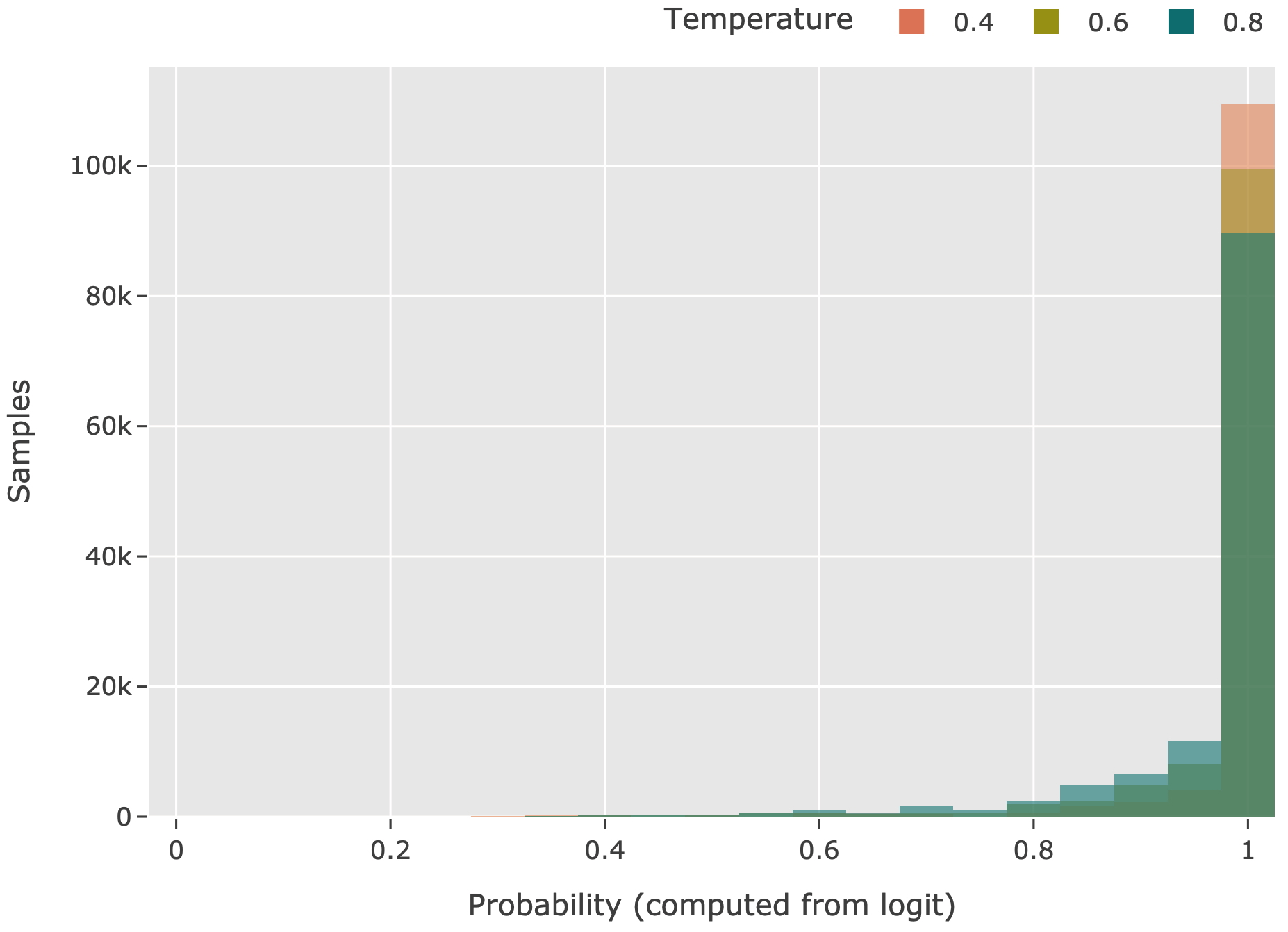}
\caption{Probability distribution of classification label tokens produced by Annotator V1. The histogram shows token probabilities computed from the model's raw logit outputs, providing insight into the model's classification confidence across the dataset. The x-axis represents probability ranges, while the y-axis shows frequency of occurrence for each probability bucket.}
\label{fig:probabilities}
\end{figure*}

Our analysis of logit outputs from the annotator LLMs demonstrates consistently high confidence in classification decisions, supporting the findings presented in the main paper. As illustrated in Figure~\ref{fig:probabilities}, the probability distribution of malware prediction tokens shows a strong skew toward values approaching $1$, indicating robust model conviction in its classifications. This concentration of high-probability predictions suggests that the model develops strong discriminative features for malware identification rather than producing uncertain or ambiguous classifications.

\subsubsection{Confidence threshold selection for filtering:}

The determination of a 90\% confidence threshold for our filtering pipeline emerged from careful analysis of the trade-off between data retention and labeling accuracy, particularly considering the known tendency of large language models towards overconfidence in their predictions.

\begin{figure*}[!t]
    \centering
    \includegraphics[width=0.65\textwidth]{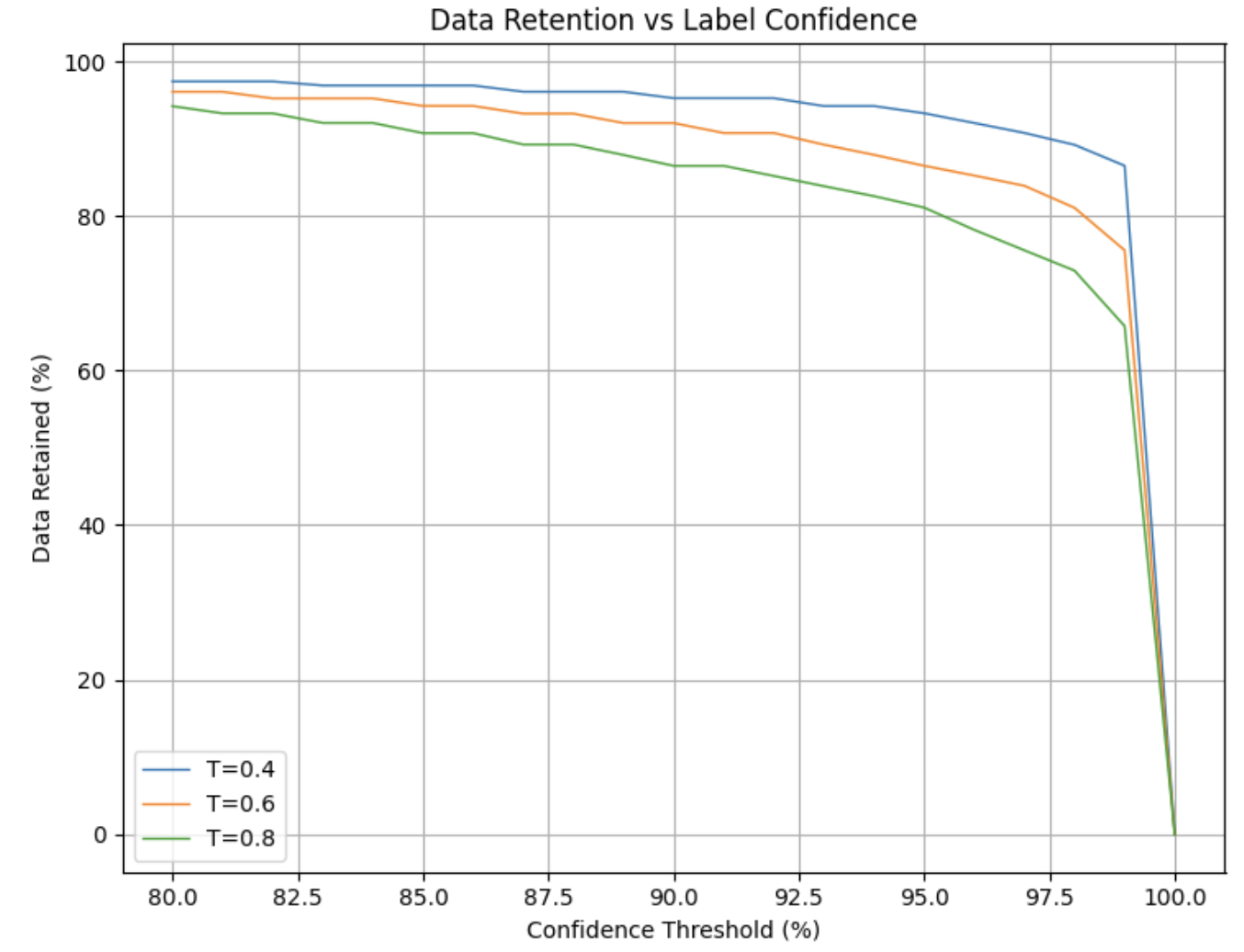}
    \caption{Impact of confidence thresholds and temperature settings on malware classification data retention. The plot demonstrates the relationship between model confidence thresholds (x-axis) and the percentage of retained samples (y-axis) across different temperature values.}
    \label{fig:label-retention-confidence}
\end{figure*}

Our label retention analysis, presented in Figure~\ref{fig:label-retention-confidence}, demonstrates that the 90\% threshold achieves optimal balance across various temperature settings, maintaining a minimum label retention rate of 86.49\%, while preserving 96.84\% of samples based on language identification. Higher confidence thresholds led to significant data loss -- up to approximately 35\% of total samples, especially at higher temperatures (e.g.~0.8) -- and, more critically, reduced the diversity of malware samples by flattening the distribution. Conversely, lower thresholds (80\% or below) proved insufficiently selective despite high data retention rates for both label and language identification, allowing too many low-quality predictions from V1 to pass through the filter.

\begin{figure*}[!t]
    \centering
    \includegraphics[width=0.65\textwidth]{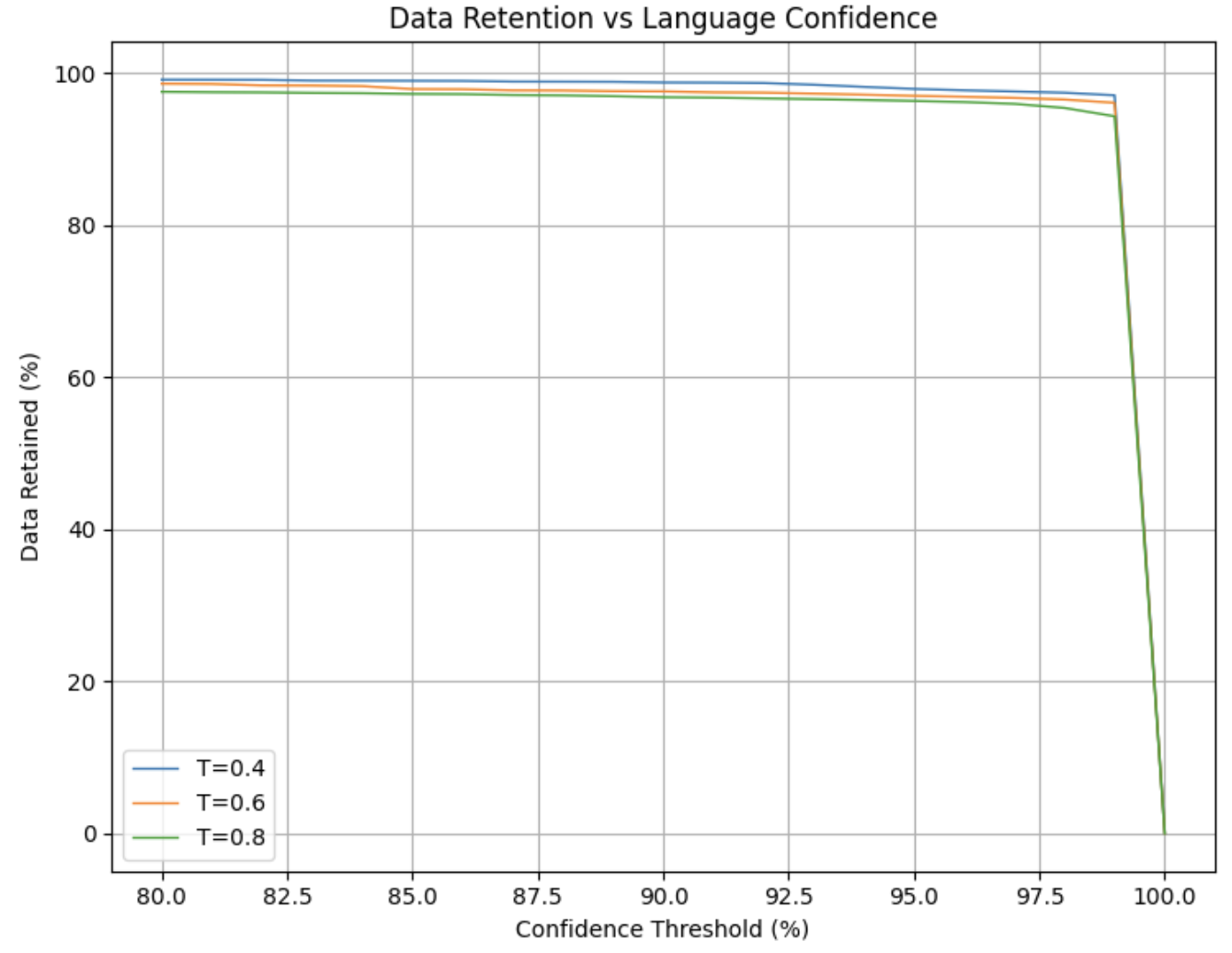}
    \caption{Impact of confidence thresholds and temperature settings on language identification data retention. The plot demonstrates the relationship between model confidence thresholds (x-axis) and the percentage of retained samples (y-axis) across different temperature values. }
    \label{fig:language-retention-confidence}
\end{figure*}

\begin{table}[tbp]
    \centering
    \begin{tabular}{cccc}
        Confidence & Temp & Label Ret. (\%) & Language Ret. (\%) \\
        \hline
        80.0\% & 0.4 & 97.40 & 99.14 \\
        80.0\% & 0.6 & 96.07 & 98.62 \\
        80.0\% & 0.8 & 94.20 & 97.55 \\
        85.0\% & 0.4 & 96.88 & 98.99 \\
        85.0\% & 0.6 & 94.23 & 97.90 \\
        85.0\% & 0.8 & 90.71 & 97.27 \\
        \textbf{90.0\%} & \textbf{0.4} & \textbf{95.23} & \textbf{98.77} \\
        \textbf{90.0\%} & \textbf{0.6} & \textbf{92.02} & \textbf{97.60} \\
        \textbf{90.0\%} & \textbf{0.8} & \textbf{86.49} & \textbf{96.84} \\
        95.0\% & 0.4 & 93.28 & 97.92 \\
        95.0\% & 0.6 & 86.49 & 97.01 \\
        95.0\% & 0.8 & 81.08 & 96.34 \\
        99.0\% & 0.4 & 86.50 & 97.09 \\
        99.0\% & 0.6 & 75.56 & 96.10 \\
        99.0\% & 0.8 & 65.76 & 94.32 \\
    \end{tabular}
    \caption{Data retention rates across different confidence thresholds and temperature settings. The table shows the percentage of samples retained based on label confidence (Label Ret.) and language identification (Language Ret.) across varying model confidence thresholds (80-99\%) and sampling temperatures (0.4-0.8). Bold entries highlight our selected operating point of 90\% confidence threshold, which balances retention rates with prediction reliability. Higher temperatures and confidence thresholds generally result in lower retention rates, particularly for label classification tasks.}
    \label{tab:confidence-retention}
\end{table}

This threshold selection reflects our strategic focus on creating high-quality training data for Annotator V2 that maintains both sample diversity and prediction accuracy. We concentrated our filtering criteria exclusively on label retention, as the base LLM demonstrated robust accuracy in programming language identification from its pre-training data (illustrated in Figure~\ref{fig:language-retention-confidence}), with malware classification remaining the primary challenge. The 90\% threshold represents an empirically optimized balance point that effectively filters unreliable predictions, while preserving sufficient dataset scale for effective iterative learning.
 
\subsection*{Qualitative Comparison with Base Model}

To establish a comprehensive baseline comparison, we conducted additional evaluations contrasting our fine-tuned annotator models against the pretrained Llama-3.3-70B-Instruct. Notably, the annotators operated with minimal instruction, receiving only a single-sentence prompt, while the base model required a more elaborate, structured prompt similar to the one used during seed data generation.

Table \ref{tab:qualitative_assessments_comparison_base_v1} presents the comparison between the base model and Annotator V1. The results show that  three of six judges prefer V1's summaries (Llama-3.3-70B-Instruct: 53.02\%, Claude-3.7-Sonnet: 52.44\% and Phi-3.5-Mini-Instruct: 55.35\%), while the remaining three show marginal preference towards the base model (ranging from 50.70\% to 51.39\%). This mixed outcome indicates that fine-tuning on the seed dataset enables competitive performance with significantly reduced prompt complexity, with V1 summaries showing modest but consistent advantages over the base model across half of the evaluators. 

\begin{table}[!t]
    \centering
    \begin{tabular}{lccc}
        \multicolumn{3}{c}{Model Evaluation Results (Base vs.~V1)} \\
        Evaluator & Base & V1 \\
        \hline
        Llama-3.3-70B-Instruct & 46.98\% & 53.02\% \\
        Claude 3.7 & 47.56\% & 52.44\%\\
        Phi-3.5-Mini-Instruct & 44.65\% & 55.35\% \\
        Mixtral-7x22B-Large & 50.70\% & 49.30\% \\
        GPT-4o & 51.16\% & 48.84\% \\
        GPT-5 & 51.39\% & 48.61\% \\
    \end{tabular}
    \caption{Comparison of summary quality between the pretrained Llama-3.3-70B-Instruct baseline and the V1 annotator (fine-tuned on seed data) across four LLM judges. Percentages represent preference rates over 215 script samples.}
    \label{tab:qualitative_assessments_comparison_base_v1}
\end{table}

Table \ref{tab:qualitative_assessments_comparison_base_v2} shows the evaluation of V2's summaries against the base model. While the percentage differences are relatively small, only two out of the six judges prefer V2 (Claude-3.7-Sonnet with 53.95\%  and Phi-3.5-Mini-Instruct with 50.70\%). This outcome is noteworthy given V2's training approach: unlike V1, which was fine-tuned on 900 high-quality, SANDBOX-informed samples, V2 was trained on approximately 101K samples generated through V1's pseudo-labeling process. The near-parity with the extensively prompted base model suggests that V2 successfully maintains summary quality despite the substantial increase in training data scale and the shift from expert-curated to model-generated annotations. This further confirms that our iterative self-training approach preserves quality while notably expanding the dataset size, although it does not yield the same preference gains observed with V1's training. This marginal variation in preference reflects the trade-off between technical precision and natural language quality also observed in the human evaluation study.

\begin{table}[!t]
    \centering
    \begin{tabular}{lccc}
        \multicolumn{3}{c}{Model Evaluation Results (Base vs.~V2)} \\
        Evaluator & Base & V2 \\
        \hline
        Llama-3.3-70B-Instruct & 50.23\% & 49.77\% \\
        Claude 3.7 & 46.05\% & 53.95\%\\
        Phi-3.5-Mini-Instruct & 49.30\% & 50.70\% \\
        Mixtral-7x22B-Large & 54.67\% & 45.33\% \\
        GPT-4o & 51.85\% & 48.15\% \\
        GPT-5 & 50.69\% & 49.31\% \\
    \end{tabular}
    \caption{Comparison of summary quality between the pretrained Llama-3.3-70B-Instruct baseline and the V2 annotator (fine-tuned on seed data) across four LLM judges. Percentages represent preference rates over 215 script samples.}
    \label{tab:qualitative_assessments_comparison_base_v2}
\end{table}

\subsection*{Methodological Details}

\subsubsection*{Fine-tuning implementation:}

Our fine-tuning approach combines cross-entropy optimization with Low-Rank Adaptation (LoRA) to efficiently adapt LLMs to our specific tasks. We implemented instruction-tuning with selective gradient computation on response tokens to optimize training efficiency. The implementation leverages the HuggingFace Transformers library~\cite{Wolf-EMNLP-2020}. The code is publicly available in our repository.

\subsubsection*{Self-training framework:}

The self-learning and synthetic data filtering processes described in the main paper are formally presented in Algorithms~\ref{alg:self-learning} and~\ref{alg:filtering}.

\vspace{0.2cm}
\noindent
\textbf{Algorithm~\ref{alg:self-learning}} outlines our iterative self-learning approach, which progressively improves the annotator model through repeated fine-tuning and filtering cycles. The algorithm takes five key inputs: a seed dataset consisting of paired inputs and outputs $(S_x, S_y)$, a set of unlabeled inputs $U_x$, an initial base LLM Annotator $A_0$, a LLM Judge $J$ for validation, and a specified number of learning iterations $k$. Starting with empty synthetic labels $U_{y,0}$, the algorithm executes $k$ iterations of three core operations: fine-tuning model $A_i$ on combined seed and previously generated synthetic data, generating new labels $\hat{U}_{y,i}$ for unlabeled inputs through inference and applying quality filtering to the generated labels, producing refined labels $U_{y,i}$. This creates a feedback loop where each iteration builds upon filtered outputs from previous generations, progressively refining the model's capabilities, while maintaining output quality through filtering. The iterative nature of this approach allows the model to leverage increasingly refined synthetic data for training, while the filtering mechanism ensures the quality of generated labels remains high throughout the process, ultimately producing the final LLM Annotator $A_k$ with enhanced performance characteristics. 

\begin{algorithm}[t]
\caption{Generic self-learning method.}\label{alg:self-learning}
\begin{algorithmic}[1]
\REQUIRE{                                 $\left(S_x, S_y\right)$\enspace Seed inputs and outputs,\\
\phantom{\textbf{Require: }}\hspace{-14pt}$U_x$\enspace Unlabeled inputs,\\
\phantom{\textbf{Require: }}\hspace{-14pt}$A_0$\enspace Base LLM,\\
\phantom{\textbf{Require: }}\hspace{-14pt}$J$\enspace LLM Judge,\\
\phantom{\textbf{Require: }}\hspace{-14pt}$k$\enspace Number of learning iterations.
}
\ENSURE{$A_k$\enspace LLM Annotator}
\STATE $U_{y,0} \gets \emptyset$
\FORALL{$i$ in ${1, 2,..., k}$}
    \STATE{$A_{i}     \gets \mathrm{FineTuning}(\left(S_x, S_y\right) \cup (U_x, U_{y,i-1}))$}
    \STATE{$\hat{U}_{y,i} \gets \mathrm{Inference}\left(A_i, U_x\right)$}
    \STATE{$U_{y,i} \gets \mathrm{Filtering}\left(\hat{U}_{y,i}, A_i, J\right)$}
\ENDFOR
\RETURN{$A_k$}
\end{algorithmic}
\end{algorithm}

\begin{algorithm}[t]
\caption{Filtering pipeline for synthetic labels.}\label{alg:filtering}
\begin{algorithmic}[1]
\REQUIRE{                                 $\hat{U}_{y}^t$\enspace Synthetic labels from temperature $t$,\\
\phantom{\textbf{Require: }}\hspace{-14pt}$T$\enspace Set of temperatures that generated $\hat{U}_{y}^t$,\\
\phantom{\textbf{Require: }}\hspace{-14pt}$A$\enspace LLM Annotator that generated $\hat{U}_y$,\\
\phantom{\textbf{Require: }}\hspace{-14pt}$J$\enspace LLM Judge,\\
\phantom{\textbf{Require: }}\hspace{-14pt}$\alpha$\enspace Confidence threshold.
}
\ENSURE{$U_y$\enspace Filtered synthetic labels}

\COMMENT{``Sanity'' check}
\STATE{$\hat{U}_{y}^t \gets \{y \mid y \in {\hat{U}_{y}^t}\enspace\text{can be parsed}\}$}

\COMMENT{Consensus check}
\STATE{$S_{\text{label}} \gets \{S \mid S\text{ SHA in }{\hat{U}_y}, |\{{y_{S,\text{label}}^{t}} \mid t \in T\}| = 1\}$}
\STATE{$\hat{U}_y^t \gets \{y_S \mid S \in S_{\text{label}}, y_S \in \hat{U}_{y}^t\}$}

\COMMENT{Confidence check}
\STATE{$\hat{U}_y^t \gets \{ y \mid y \in \hat{U}_y^t, \text{logit } l \text{ for }y_{\text{label}}, e^{l} \geq \alpha \}$}

\COMMENT{Coherence check}
\STATE{$\hat{U}_y^t \gets \{ y \mid y \in \hat{U}_y^t, J(y_{\text{summary}}) = y_{\text{label}} \} $}

\COMMENT{``Sampling''. We set the temperature to $0.6$.}
\STATE{$U_y \gets \hat{U}_y^{0.6}$}

\RETURN{$U_y$}

\end{algorithmic}
\end{algorithm}

\begin{figure*}[!t]
    \centering
    \includegraphics[width=0.75\textwidth]{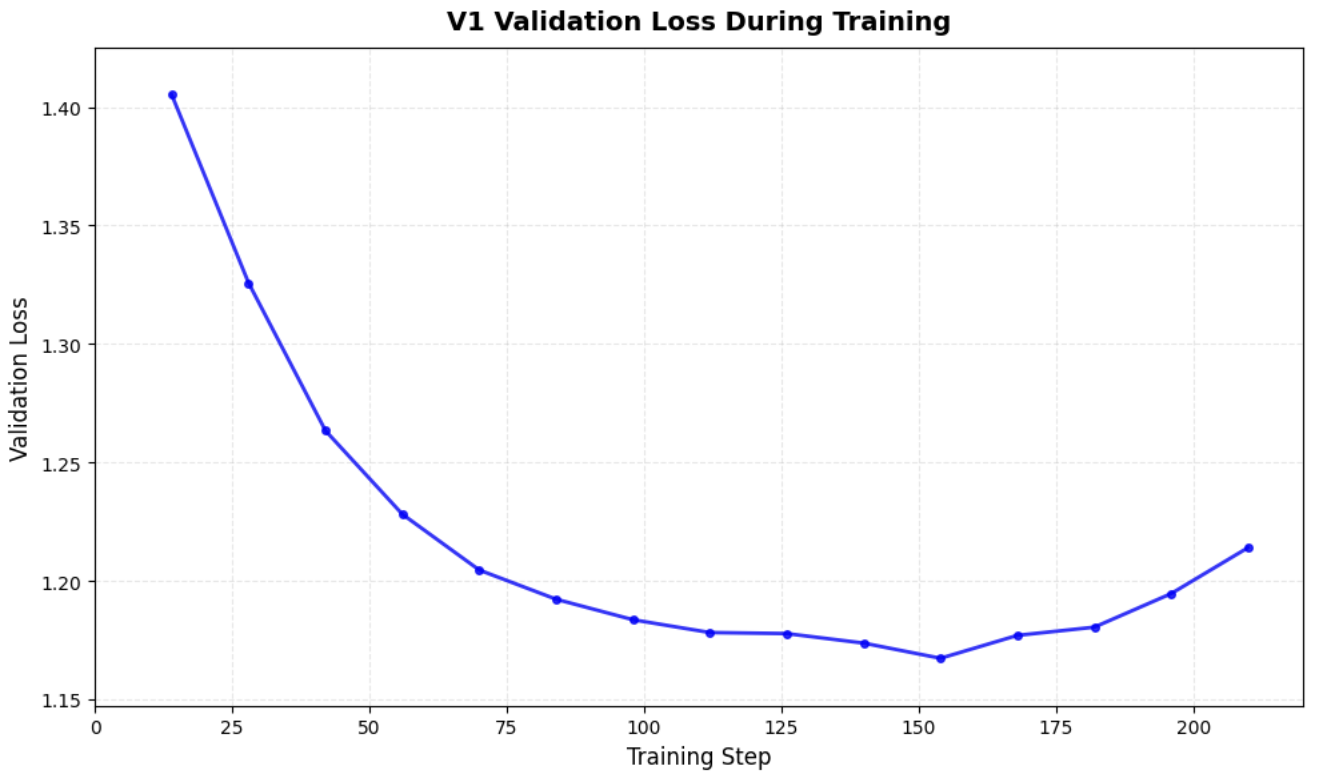}
    \caption{The curve demonstrates smooth convergence with optimal performance achieved at epoch 11 (step 154), before signs of overfitting emerge.}
    \label{fig:v1-loss}
\end{figure*}

\vspace{0.2cm}
\noindent
\textbf{Algorithm~\ref{alg:filtering}} implements our multi-stage filtering pipeline that ensures quality control over synthetic labels generated by LLM Annotators. The algorithm accepts synthetic labels $\hat{U}_y^t$ generated at different temperatures $T$, along with the source LLM Annotator $A$, an LLM Judge $J$ for validation and a confidence threshold $\alpha$. The filtering process consists of five sequential stages. First, a basic sanity check ensures all labels are syntactically parsable. Second, a consensus check identifies samples where the label predictions remain consistent across different temperature settings, retaining only those with unanimous label assignments. Third, a confidence check filters labels based on the model's logit scores, keeping only those exceeding the threshold $\alpha$. Fourth, a coherence check validates labels through the LLM Judge, ensuring alignment between the summary content and assigned label. Finally, the sampling stage selects outputs generated at temperature 0.6, producing the final filtered set $U_y$.
This sequential filtering progressively refines the label set through set comprehension operations, ensuring only high-quality, consistent and coherent labels survive for subsequent model training.

The presented training framework maintains flexibility for adaptation to different contexts. The algorithms support multiple modification paths. Temperature settings can be replaced with alternative diversity-controlling hyperparameters. Sampling strategies during synthetic label generation can be adjusted to meet specific needs. Filtering criteria can be modified to accommodate varying quality requirements. This adaptability ensures the framework's applicability across diverse use cases while maintaining its core self-learning functionality.

\begin{figure*}[!t]
    \centering
    \includegraphics[width=0.75\textwidth]{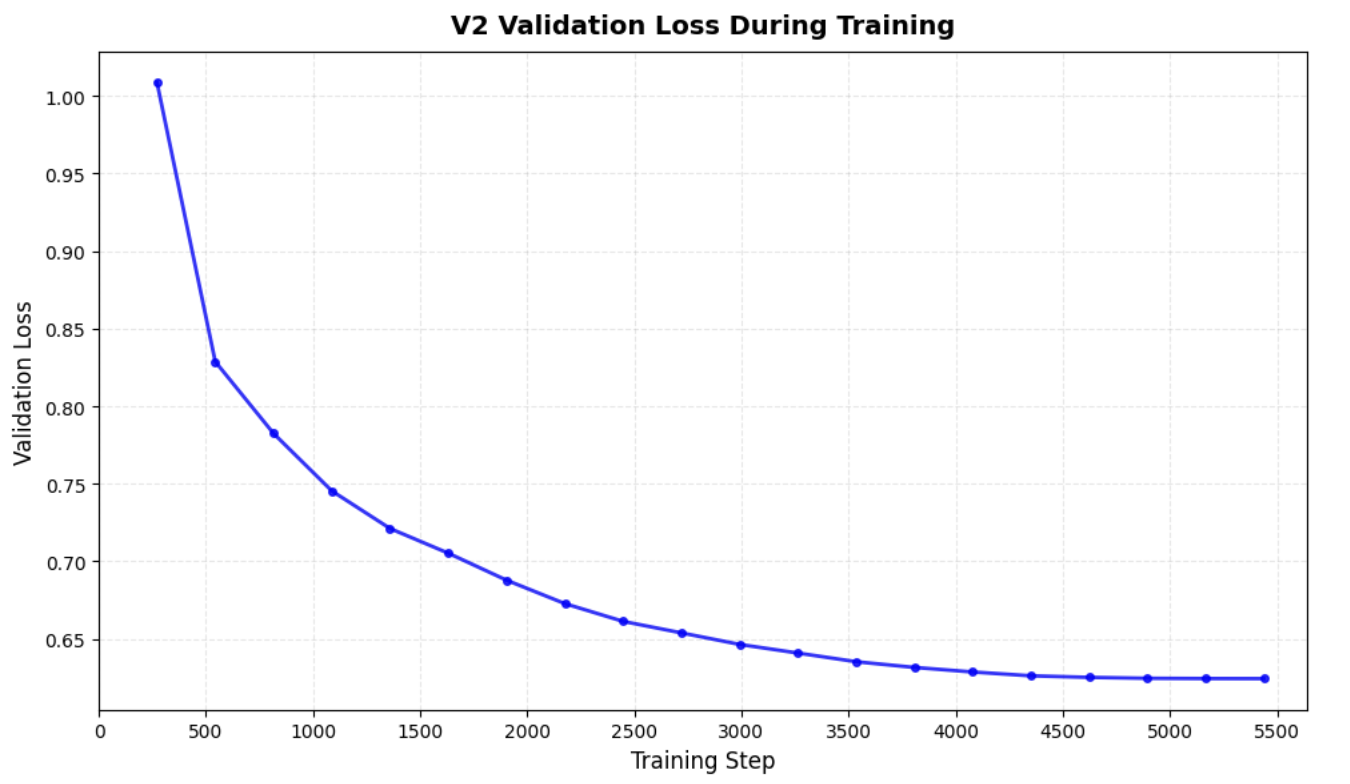}
    \caption{Validation loss progression for Annotator V2, showing extended training required for learning from the augmented synthetic corpus. Optimal performance reached at epoch 13.}
    \label{fig:v2-loss}
\end{figure*}

\subsection*{Experimental Details}

\subsubsection*{Computational infrastructure and resource utilization:}
Our experiments were conducted on a cluster comprising multiple nodes, each equipped with 8 NVIDIA H100 Mega GPUs, managed through the Slurm scheduling system. The computational requirements varied between model versions. Annotator V1 training utilized 2 nodes (16 H100 GPUs), completing 15 epochs in approximately 1.5 hours. In contrast, Annotator V2 required 20 nodes (160 H100 GPUs) for training, completing 20 epochs over approximately 2 days, due to the substantially larger corpus size.

The inference process proved computationally intensive, requiring approximately one day to process predictions for 157,000 samples when distributed across 20-24 nodes (160-192 H100 GPUs). This extended processing time stems from both model complexity and substantial data handling requirements, with dataset loading creating notable computational bottlenecks. Given these significant computational demands, we made the pragmatic decision to omit hyperparameter tuning, focusing our resources on ensuring robust primary experiments.

\begin{figure*}[!t]
    \centering
    \includegraphics[width=0.75\textwidth]{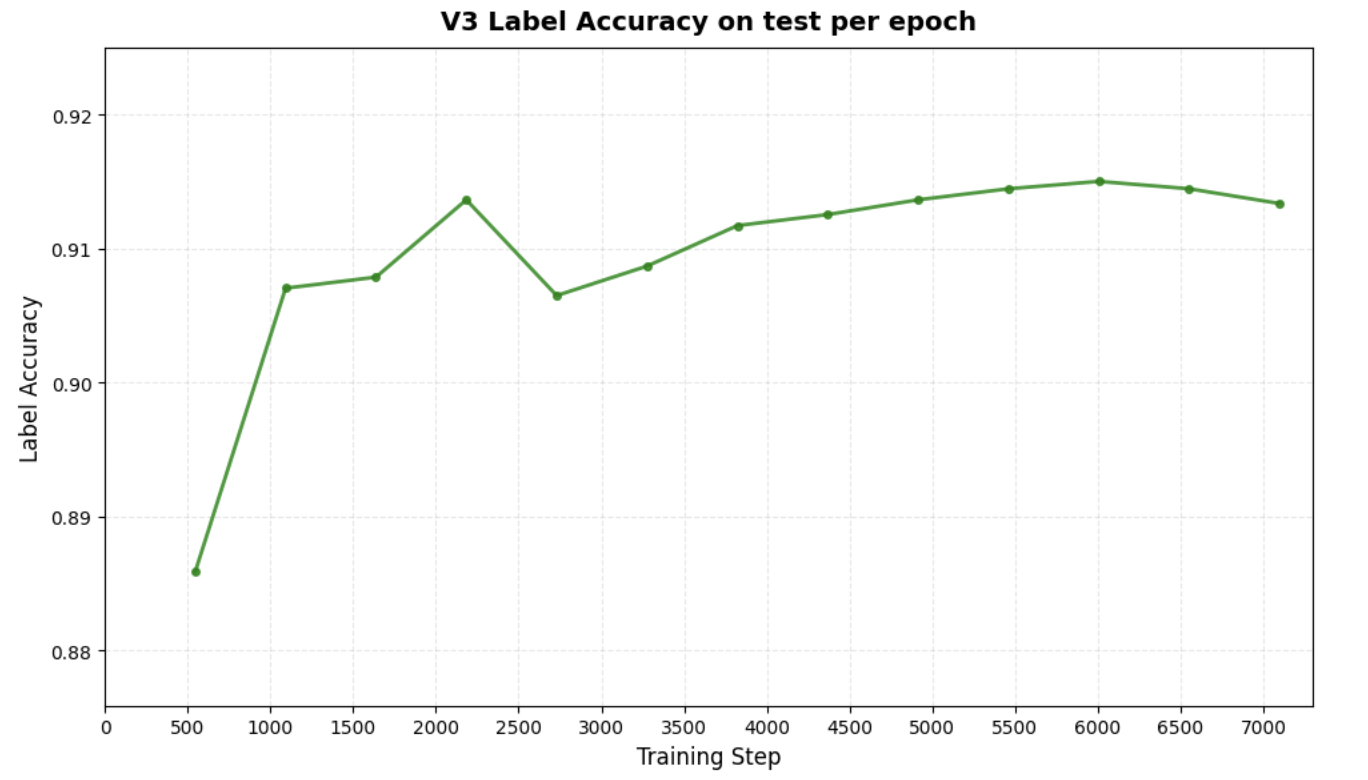}
    \caption{Performance evaluation of Annotator V3's labeling accuracy across training epochs, with peak accuracy of 91.53\% at epoch 10, matching V2's performance.}
    \label{fig:v3-label}
\end{figure*}

\begin{figure*}[!t]
    \centering
    \includegraphics[width=0.75\textwidth]{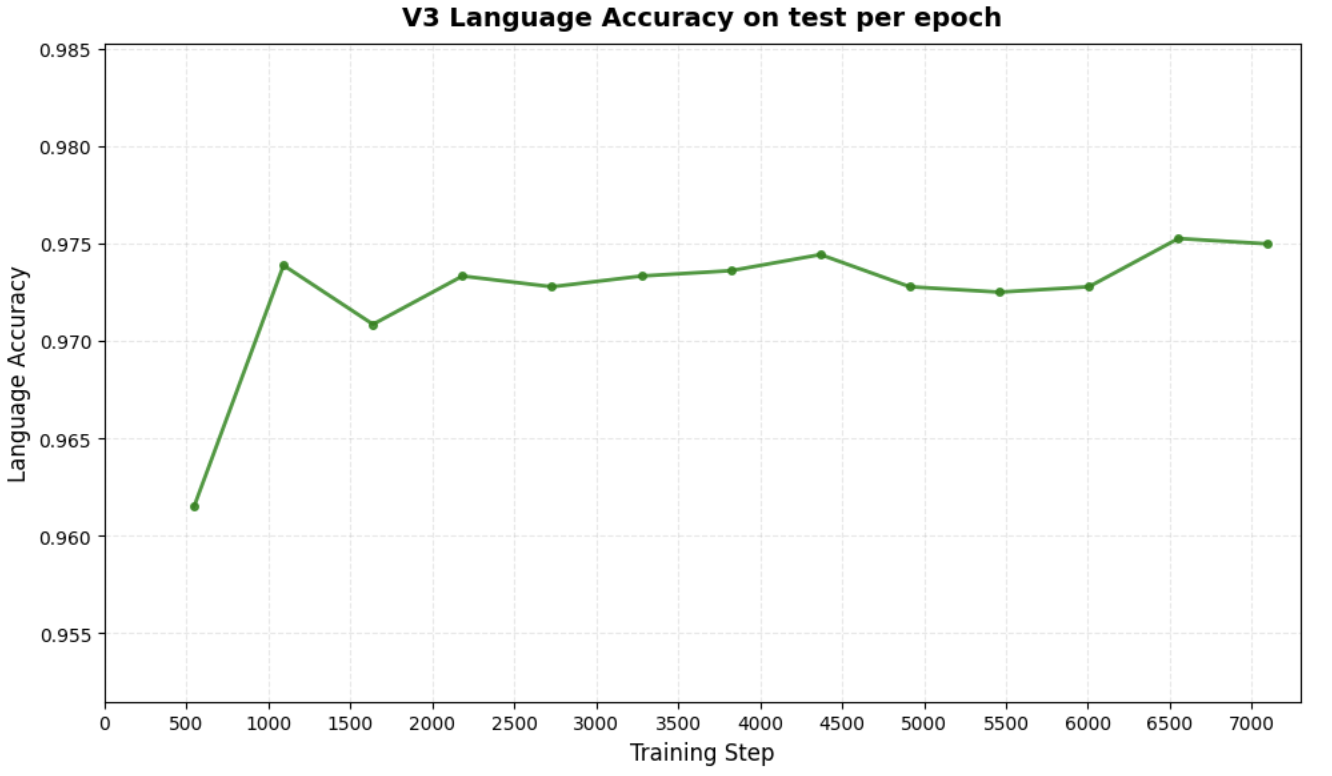}
    \caption{Language identification accuracy progression for Annotator V3, demonstrating peak performance of 97.52\% at epoch 13, marginally surpassing V2's 97.14\% accuracy.}
    \label{fig:v3-language}
\end{figure*}

\subsubsection*{Evaluation Methodology}

\paragraph*{Performance metrics:} 
We evaluated model performance using standard classification metrics. Label and language accuracy were computed as the proportion of correctly classified samples:
\begin{equation}
\text{Accuracy} = \frac{\text{TP} + \text{TN}}{\text{TP} + \text{TN} + \text{FP} + \text{FN}}.
\end{equation}

For comparative evaluation using LLM judges, we calculated the win rate as:
\begin{equation}
\text{Win Rate} = \frac{\text{Number of Wins}}{\text{Total Pairwise Comparisons}}.
\end{equation}

Statistical significance was assessed using McNemar's test~\citep{McNemar-Psychometrika-1947}:
\begin{equation}
\chi^2 = \frac{(b-c)^2}{b+c},
\end{equation}
where $b=110$ samples were predicted as malicious by V1, but benign by V2, and $c=49$ vice versa, yielding $\chi^2=23.402$ ($p<10^{-5}$), rejecting the null hypothesis that $p_b = p_c$.

\paragraph*{Exploration of a third generation model (V3):}

Our investigation of a third-generation annotator (V3) revealed diminishing returns. During the annotation of an expanded dataset ($\approx$250K samples), V2 exhibited increased prediction noise compared to V1, necessitating modified filtering strategies, including the exclusion of a temperature of 0.8 in our consistency checks, to prevent losing approximately 50\% of the generated samples due to excessive filtering by temperature-induced mismatches. Despite these adjustments and training V3 on the expanded corpus, performance metrics showed marginal improvements at best, as seen in Figures \ref{fig:v3-label} and \ref{fig:v3-language}: label accuracy peaked at 91.53\% (epoch 10), matching V2's performance, while language accuracy reached 97.52\% (epoch 13), marginally exceeding V2's 97.14\%. These results suggest optimal performance plateaus at two iterations.

\paragraph*{Training dynamics and model selection:}

The graphs depicted in Figure~\ref{fig:v1-loss} and \ref{fig:v2-loss}) represent the validation losses during the training phases for V1 and V2 models. They showcase slightly different convergence patterns that guided our checkpoint selection process. For Annotator V1 (Figure~\ref{fig:v1-loss}), trained on the 900 high-quality seed samples, the validation loss decreases smoothly throughout training. Optimal performance is reached at epoch 11 (step 154), where the loss curve stabilizes before showing signs of overfitting in subsequent epochs.

In contrast, Annotator V2 (Figure \ref{fig:v2-loss}), trained on the expanded dataset of seed data plus 100K V1-generated samples, required extended training until epoch 13 (step 5356). The training curve exhibits the same initial steep decline  as V1, followed by gradual improvements, reflecting the complexity of learning from the augmented synthetic corpus. Both models were trained until validation loss convergence, with checkpoints selected at minimum loss to ensure robust performance on unseen data.

\end{document}